\documentclass{plain_class}
\jno{dqnxxx}
\usepackage[round]{natbib}
\usepackage{graphicx}
\usepackage{amsmath,amsthm,bm,mathrsfs}
 \newtheoremstyle{theorem}{6pt}{6pt}{\rm}{}{\sffamily}{ }{ }{}
 \theoremstyle{theorem}

 \newtheoremstyle{algorithm}{6pt}{6pt}{\rm}{}{\sffamily}{ }{ }{}
 \theoremstyle{algorithm}

 \newtheoremstyle{lemma}{6pt}{6pt}{\rm}{}{\sffamily}{ }{ }{}
 \theoremstyle{lemma}

\newtheoremstyle{case}{6pt}{6pt}{\rm}{}{\sffamily}{. }{ }{}
 \theoremstyle{case}

 \newtheoremstyle{statement}{6pt}{6pt}{\rm}{}{\sffamily}{ }{ }{}
\theoremstyle{statement}

 \newtheoremstyle{corollary}{6pt}{6pt}{\rm}{}{\sffamily}{ }{ }{}
 \theoremstyle{corollary}

  \newtheoremstyle{definition}{6pt}{6pt}{\rm}{}{\sffamily}{ }{ }{}
 \theoremstyle{definition}

\newtheoremstyle{example}{6pt}{6pt}{\rm}{}{\sffamily}{ }{ }{}
\theoremstyle{example}

\newtheoremstyle{remark}{6pt}{6pt}{\rm}{}{\sffamily}{ }{ }{}
\theoremstyle{remark}

\newtheoremstyle{approximation}{6pt}{6pt}{\rm}{}{\sffamily}{ }{ }{}
\theoremstyle{approximation}

\newtheoremstyle{scheme}{6pt}{6pt}{\rm}{}{\sffamily}{ }{ }{}
\theoremstyle{scheme}

\newtheoremstyle{Algorithm}{6pt}{6pt}{\rm}{}{\sffamily}{ }{ }{}
\theoremstyle{Algorithm}

\newtheoremstyle{Assumption}{6pt}{6pt}{\rm}{}{\sffamily}{ }{ }{}
\theoremstyle{Assumption}

\newtheoremstyle{proposition}{6pt}{6pt}{\rm}{}{\sffamily}{ }{ }{}
\theoremstyle{proposition}

\newtheoremstyle{hypo}{6pt}{6pt}{\rm}{}{\sffamily}{ }{ }{}
 \theoremstyle{hypo}

  \newtheoremstyle{Step}{6pt}{6pt}{\rm}{}{}{ }{ }{}
 \theoremstyle{Step}

\numberwithin{equation}{section}

\usepackage[latin1]{inputenc}
\usepackage{amsfonts}
\usepackage{amssymb}
\usepackage{paralist}
\usepackage{esdiff}
\usepackage{amsopn}
\usepackage{titlesec}
\usepackage{listings}
\usepackage{fancyref}
\usepackage{hyperref}
\usepackage{xcolor}

\def\barroman#1{\sbox0{#1}\dimen0=\dimexpr\wd0+1pt\relax
  \makebox[\dimen0]{\rlap{\vrule width\dimen0 height 0.06ex depth 0.06ex}%
    \rlap{\vrule width\dimen0 height\dimexpr\ht0+0.03ex\relax 
            depth\dimexpr-\ht0+0.09ex\relax}%
    \kern.5pt#1\kern.5pt}}

\titleformat{\subsection}{\normalsize\bfseries}{}{0pt}{}

\begin{document}

\title{{\color{black}Relating cell shape and mechanical stress in a spatially disordered \\ epithelium using a vertex-based model}}
\author{ {\sc Alexander Nestor-Bergmann,$^{1,2}$ Georgina Goddard,$^2$ \\ Sarah Woolner$^2$ \& Oliver E. Jensen$^1$}\\[2pt]
$^1$ School of Mathematics, University of Manchester, 
 Manchester M13 9PL, UK \\
$^2$ Wellcome Trust Centre for Cell-Matrix Research,  Faculty of Biology, Medicine and Health, University of Manchester, Manchester M13 9PT UK \\[6pt]
{\rm [Received on \today]}\vspace*{6pt}}
\pagestyle{headings}
\markboth{NESTOR-BERGMANN, GODDARD, WOOLNER \& JENSEN}{\rm RELATING SHAPE AND STRESS IN  EPITHELIUM}
\maketitle

\begin{abstract}
{Using a popular vertex-based model to describe a spatially disordered planar epithelial monolayer, we examine the relationship between cell shape and mechanical stress at the cell and tissue level.   Deriving expressions for stress tensors starting from an energetic formulation of the model, we show that the principal axes of stress for an individual cell align with the principal axes of shape, and we determine the bulk effective tissue pressure when the monolayer is isotropic at the tissue level.  Using simulations for a monolayer that is not under peripheral stress, we fit parameters of the model to experimental data for \textit{Xenopus} embryonic tissue.  The model predicts that mechanical interactions can generate mesoscopic patterns within the monolayer that exhibit long-range correlations in cell shape.  The model also suggests that the orientation of mechanical and geometric cues for processes such as cell division are likely to be strongly correlated in real epithelia.  {\color{black} Some limitations of the model in capturing geometric features of \textit{Xenopus} epithelial cells are highlighted.}}
\end{abstract}

\section{Introduction}

Many essential aspects of cell behaviour are controlled, both directly and indirectly, by mechanical cues \citep{Huang:1999, wozniak2009}.  For example, cell density and substrate adhesion have been shown to affect cell proliferation \citep{huang2000, Streichan:2014}, while cell division orientation appears to be regulated by mechanical feedback \citep{thery:2006, Minc:2011, Fink:2011, Wyatt2015}.  Many morphogenetic processes, such as gastrulation and convergent extension \citep{martin2009}, are mechanical processes inducing significant changes to the stresses within the tissue \citep{Lecuit:2007}.  However, despite its significance in development, the mechanical state of tissues remains poorly characterised in comparison to some aspects of genetics and biochemical signalling. 

The geometric properties of cells are governed by cell adhesions and cytoskeletal mechanics \citep{Kafer:2007, kiehart2000}, which in turn feed into global tissue dynamics \citep{guillot2013, martin2009, shraiman2005}.  The mechanical state of an individual cell is largely dependent on its interaction with its neighbours and adhesion to the extracellular matrix.  Experimental techniques such as laser ablation \citep{campinho2013, hutson2003forces, mao2013} and atomic force microscopy (AFM) \citep{hoh1994surface} have been used to characterise cell mechanics; laser ablation reveals cell-level forces by making small slices in the tissue and observing the recoil velocity of cells, while AFM attempts to deduce the local mechanical properties of a tissue by performing small indentations using a mechanical cantilever.  While revealing, such experimental  techniques are invasive and typically require modelling for the interpretation of measurements.  Live fluorescent imaging combined with high resolution microscopy offers alternative insights into developmental processes such as gastrulation \citep{rauzi2008nature, Heller2016103}.  Measurements of cell shape over time allows inference of mechanical stress \citep{chiou2012mechanical, ishihara2012, xu2015changes, xu2016oriented}, based on an underlying mathematical model.  This non-invasive approach has led to significant growth in mathematical modelling of epithelial cell mechanics {\color{black} in two and three dimensions \citep{bielmeier2016, brodland2010video, collinet2015, hannezo2014, hilgenfeldt2008, okuda2013, sugimura2016, tetley2016}}.  However without direct measurements of stress, mechanical predictions taken from geometric data alone are only as good as the constitutive models from which the predictions are derived.

Theoretical models of epithelial mechanics fall into a number of classes, including cellular Potts \citep{Graner:1992}, cell-centre \citep{osborne2010}, vertex-based \citep{Farhadifar:2007, fletcher2014vertex, nagai2001dynamic, Staple:2010} and continuum models \citep{edwards2007, nelson2011}. Vertex-based models exploit the polygonal shape commonly adopted by tight-packed cells in a monolayer, characterising the monolayer as a network of cell edges meeting (typically) at trijunctions.    {\color{black}Typically}, vertices are assumed to move down gradients of a mechanical energy, often subject to a viscous drag; the network topology changes intermittently as cells intercalate, divide or are extruded.  It is of interest to relate such cell-level models, describing cells as individual entities that can evolve at discrete time intervals, to continuum models describing the smooth changes of a tissue in space and time.  Some progress has been made in upscaling spatially periodic cell distributions in one \citep{fozard2010} and two dimensions \citep{Murisic:2015} using homogenization approaches, {\color{black} or by direct coarse-graining \citep{ishihara2016}}.  Simulations have revealed {\color{black}striking} properties of more realistic disordered networks in two dimensions \citep{bi2015, Staple:2010}, such as a rigidity transition characteristic of a glassy material.   Abundant imaging data makes parameter estimation feasible, allowing models to be tested quantitatively and used to explore new biological hypotheses.

In this paper, working in the framework of a {\color{black}popular} vertex-based model describing a planar monolayer of {\color{black}mechanically (but not geometrically)} identical cells, we derive expressions for the stress tensor at the cell and tissue level, and use these results to understand the relationship between a cell's shape and its mechanical environment, showing that the principal axes of the cell's stress and shape tensors align.   We parameter-fit simulations to images of {\color{black}\textit{Xenopus}} embryonic epithelia, using cell {\color{black}area} over polygonal classes as a measure.  Of particular interest is the manner in which mechanical effects constrain the spatial disorder that is intrinsic to epithelial monolayers, which we characterise using simulations, highlighting the appearance of spatial patterns reminiscent of force chains in granular materials.  We also discuss the role of the stress acting on the monolayer's periphery in determining the size and shape of cells.  

\section{Experiments}
\label{sec:experiments}

\begin{figure}
	\centering
	\includegraphics[width=\textwidth]{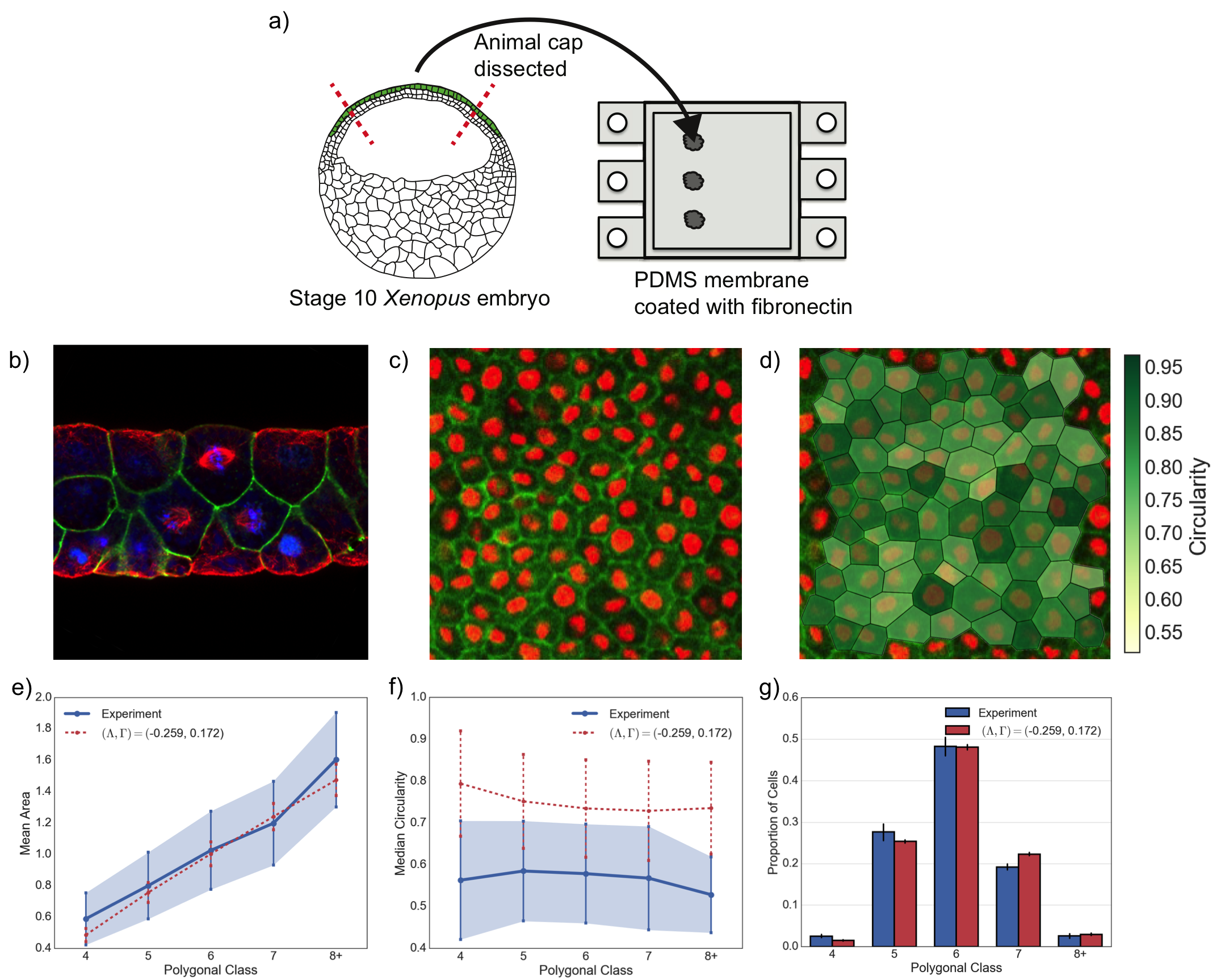}
	\caption{Experimental setup and data analysis. (a) Animal cap tissue was dissected from stage-10 \textit{Xenopus laevis} embryos and cultured on PDMS membrane. (b) Side-view confocal image of the animal cap (top:apical; bottom:basal), stained for microtubules (red), beta-catenin (green) and DNA (blue). A mitotic spindle is visible in the centremost apical cell. The animal cap is a multi-layered epithelial tissue; we analyse just the outer, apical, cell layer. (c) The apical cell layer of the animal cap tissue is imaged live using confocal microscopy (green, GFP-$\alpha$-tubulin; red, cherry-histone2B).  (d) The cell edges are manually traced and cell shapes are derived computationally\textcolor{black}{, being polygonised using the positions of cell junctions.  (e) Mean normalised area as a function of polygonal class showing mean and one standard deviation, from experiments (solid and shaded) and simulation (dashed) with parameters $\Lambda$, $\Gamma$ as shown with $P_{\mathrm{ext}}=0$. Cell areas were normalised relative to the mean of each experiment. (f) Circularity as a function of polygonal class showing mean and one standard deviation, from experiments (solid and shaded) and simulation (dashed) using the same parameters as in (e).  (g)  Proportions of total cells in each polygonal class in experiments (left bar) and simulations (right bar). Error bars represent $95\%$ confidence intervals calculated from bootstrapping the data.  }}
    \label{fig:experiments}
\end{figure}

Experimental data were collected using tissue from the albino \textit{Xenopus laevis} frog embryo.  Animal cap tissue was dissected from the embryo at stage 10 of development (early gastrula stage) and cultured on a  $20\text{mm} \times 20\text{mm} \times 1\text{mm}$, fibronectin-coated, elastomeric PDMS substrate (Figure~\ref{fig:experiments}a).  The animal cap tissue is a multi-layered (2-3 cells thick) epithelium (Figure~\ref{fig:experiments}b), which maintains its \textit{in vivo} structure when cultured externally for the time period of our experiments (up to five hours). This system has the advantage of closely resembling \textit{in vivo} tissue whilst also giving the ability to control peripheral stress on the tissue. For this work, a 0.5mm uniaxial stretch was applied to the PDMS substrate, which ensured that it did not buckle under gravity or the weight of the animal cap. This small stretch was found to have no measurable effect on cell geometry (data not shown) and we therefore assume that there is negligible peripheral stress on the tissue. The apical cell layer of the animal cap tissue was imaged using a Leica TCS SP5 AOBS upright confocal microscope (Figure~\ref{fig:experiments}c) and cell boundaries were segmented manually (Figure~\ref{fig:experiments}d), {\color{black}representing each cell as a polygon with vertices coincident with those in images.  The vast majority of vertices were classifiable as trijunctions.}

Letting a cell, $\alpha$, have {\color{black}$Z_{\alpha}$ vertices} defining its boundary, we characterise the shape of the cell using {\color{black}its area $\tilde{A}_\alpha$} and shape tensor, {\color{black}$\tilde{\mathsf{S}}_{\alpha}$}, defined {\color{black}with respect to cell vertices} as
\begin{equation}
    \label{eq:shape_tensor}
{\color{black}  
	\tilde{A}_{\alpha}=\sum_{i=0}^{Z_{\alpha}-1} \tfrac{1}{2} \hat{\mathbf{z}} \cdot (\tilde{\mathbf{R}}_{\alpha}^i\times\tilde{\mathbf{R}}_{\alpha}^{i+1}),\quad
  \tilde{\mathsf{S}}_{\alpha} =\frac{1}{Z_\alpha} \sum_{i=0}^{Z_{\alpha}-1} \tilde{\mathbf{R}}_{\alpha}^i \otimes \tilde{\mathbf{R}}_{\alpha}^i, }
\end{equation}
where {\color{black}$\tilde{\mathbf{R}}_{\alpha}^i$} is the vector running from the cell centroid to {\color{black}vertex} $i$ {\color{black}and $\hat{\mathbf{z}}$ is a unit vector pointing out of the plane}.  $\tilde{\mathsf{S}}_{\alpha}$ has eigenvalues $(\lambda_{\alpha 1}, \lambda_{\alpha 2})$ with $\lambda_{\alpha 1}\geq \lambda_{\alpha_2}>0$. The eigenvector associated with the larger (smaller) eigenvector defines the major (minor) principal axis of cell shape, the two axes being orthogonal.  The {\color{black}circularity parameter $C_{\alpha} = {\lambda_{\alpha 2}/\lambda_{\alpha 1}}\in(0,1]$} indicates how round a cell is.   

The variation {\color{black}of cell area and circularity} across an individual monolayer is illustrated in Figure~\ref{fig:experiments}(e,f), {\color{black}distributed across} the cells' polygonal class $Z_\alpha$ (number of neighbours).   {\color{black}The distribution of cell number across polygonal class is shown in Figure~\ref{fig:experiments}(g).}  The majority of cells have between 5 and 7 neighbours; we observed no 3-sided cells.  {\color{black}The mean area per polygonal class across all experiments, normalised to the mean of the population from each experiment, was $\mathbf{A}^{\;\text{exp}} = \{ \bar{A}_{4}^{\;\text{exp}}, \bar{A}_{5}^{\;\text{exp}}, \bar{A}_{6}^{\;\text{exp}}, \bar{A}_{7}^{\;\text{exp}}, \bar{A}_{8+}^{\;\text{exp}} \} = \{ 0.59, 0.80, 1.03, 1.20, 1.60 \}$ (Figure \ref{fig:experiments}e).  $\bar{A}_{8+}^{\;\text{exp}}$ represents the mean area of cells with 8 or more sides.   Similarly, the average circularity per polygonal class across all experiments, $\bar{C}_{i}^{\;\text{exp}}$, was $\mathbf{C}^{\;\text{exp}} = \{ 0.56,$ $0.58, 0.58, 0.57, 0.53 \}$ (Figure \ref{fig:experiments}f).   As explained below, we used $\mathbf{A}^{\;\text{exp}}$ to fit parameters of the vertex-based model (Figure \ref{fig:experiments}e).}

\section{The vertex-based model} 
\label{sec:development_of_the_model}

In this section we derive expressions for cell and tissue stress using the vertex-based model and describe our simulation {\color{black}methodology}.   We explain relationships between cell stress and cell shape and discuss the mechanical properties of the monolayer.  

\subsection{3.1 Geometry of the monolayer network}

We represent an epithelial monolayer as a planar network of $N_{v}$ vertices, labelled $j=1,\dots, N_{v}$, connected by straight edges and bounding $N_{c}$ polygonal cells, labelled $\alpha=1,\dots,N_{c}$.  The vector from the coordinate origin to vertex $j$ is given by $\tilde{\mathbf{R}}^{j}(\tilde{t})$; here tildes denote dimensional variables and $\tilde{t}$ is time. Quantities specific to cell $\alpha$ are defined relative to its centroid $\tilde{\mathbf{R}}_\alpha$.  Cell $\alpha$ has $Z_{\alpha}$ vertices labelled anticlockwise by $i = 0,1,2,\dots, Z_{\alpha}-1$ relative to $\tilde{\mathbf{R}}_\alpha$. We define $\tilde{\mathbf{R}}_{\alpha}^{i}$ as the vector from the cell centroid to vertex $i$, such that $\sum_{i=0}^{Z_{\alpha}-1} \tilde{\mathbf{R}}_{\alpha}^i=\mathbf{0}$. Anticlockwise tangents are defined by $\tilde{\mathbf{t}}_{\alpha}^{i} = \tilde{\mathbf{R}}_{\alpha}^{i+1} - \tilde{\mathbf{R}}_{\alpha}^{i}$, unit vectors along a cell edge by $\hat{\mathbf{t}}_{\alpha}^i$ and outward normals to edges by $\tilde{\mathbf{n}}_{\alpha}^i=\tilde{\mathbf{t}}_{\alpha}^{i}\times \hat{\mathbf{z}}$. 
The length $\tilde{l}_{\alpha}^{i}$ of an edge belonging to cell $\alpha$ between vertices $i$ and $i+1$, {\color{black}and the cell perimeter $\tilde L_{\alpha}$, are given by}
\begin{equation}
	\tilde{l}_{\alpha}^{i} = \left(\tilde{\mathbf{t}}_{\alpha}^i\cdot \tilde{\mathbf{t}}_{\alpha}^i \right)^{1/2},\quad
		\tilde{L}_{\alpha}=\sum_{i=0}^{Z_{\alpha}-1} \tilde{l}_{\alpha}^{i}.
\end{equation}
The cell area (assuming convex polygons), $\tilde{A}_{\alpha}$, and shape tensor, {\color{black}$\tilde{\mathsf{S}}_\alpha$, are given by (\ref{eq:shape_tensor})}.

\begin{figure}
	\centering
	\includegraphics[width=0.6\textwidth]{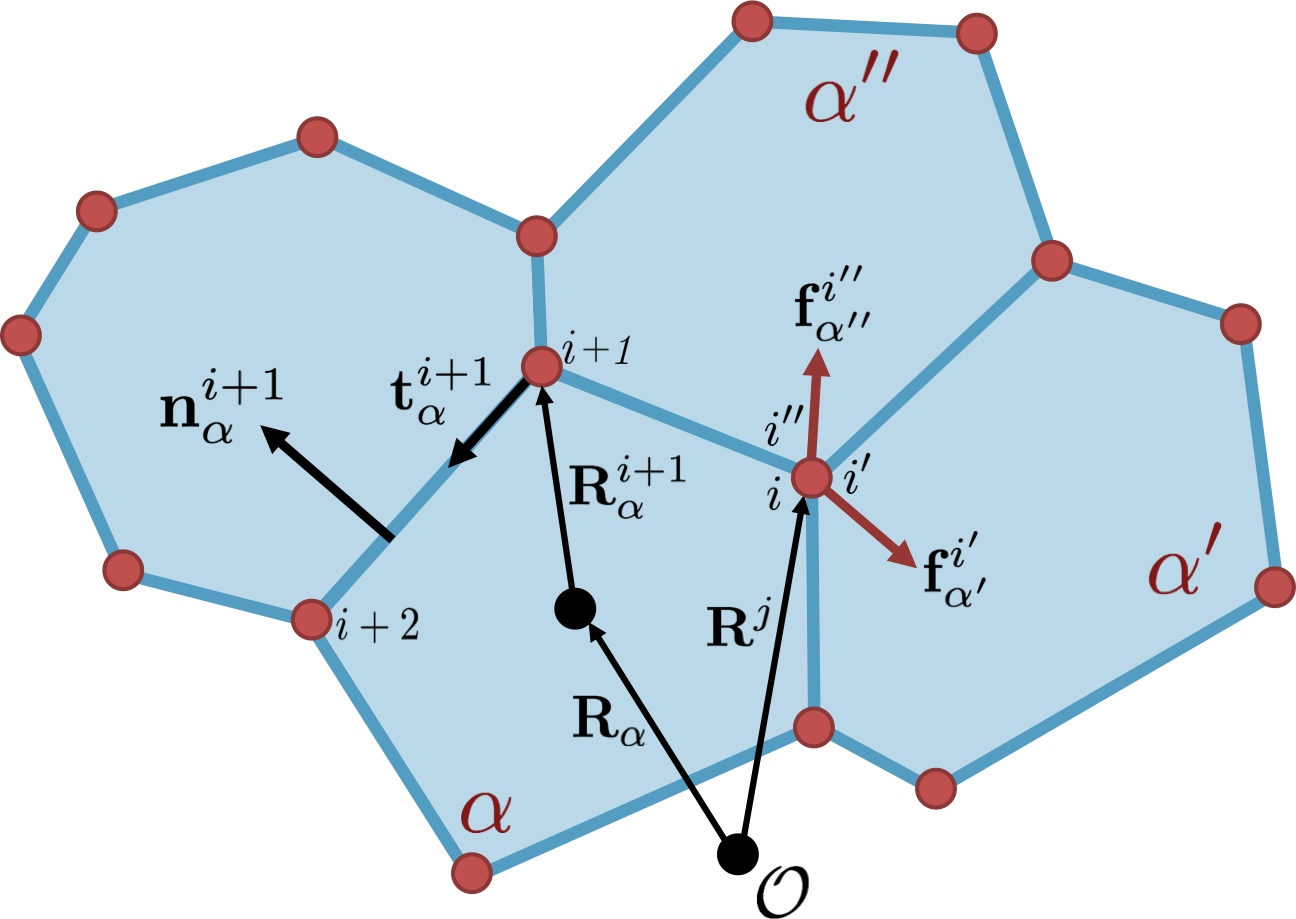}
	\caption{Representation of \textcolor{black}{disordered} cell geometry.  Cell $\alpha$ has its centroid at $\mathbf{R}_{\alpha}$ relative to a fixed origin, $\mathcal{O}$.  The position of vertex $i$ of cell $\alpha$ is given equivalently via $\mathbf{R}^{i}_{\alpha}$, relative to the centroid, or $\mathbf{R}^{j}$, relative to $\mathcal{O}$. For a vertex (trijunction) at $\mathbf{R}^{j}$, there exist three vectors, $\mathbf{R}^{i}_{\alpha}, \mathbf{R}^{i^{\prime}}_{\alpha^{\prime}}, \mathbf{R}^{i^{\prime\prime}}_{\alpha^{\prime\prime}}$ for cells, $\alpha, \alpha^{\prime}, \alpha^{\prime\prime}$, pointing to the same vertex. Cell properties, such as area and tangents along edges, are defined relative to the cell centroid.}
    \label{fig:geometry}
\end{figure}

Vectors defined relative to a cell centroid are labelled by a greek subscript, $\alpha$; vertices belonging to the cell have latin superscripts, $i$, i.e. $\tilde{\mathbf{R}}^{i}_{\alpha}$. Vectors without a greek subscript are defined relative to the coordinate origin, and have unique latin superscripts, $j$ i.e. $\tilde{\mathbf{R}}^{j}$. The matrix capturing the mapping from the vertex labels, $j$, to the vertex labels, $i$, within every cell, $\alpha$, is defined as
\begin{equation}
    c_{\alpha}^{ij} = \begin{cases}  1 \qquad \text{if } \exists \; j\prime \in \{1,\dots, N_{v}\} \mid \tilde{\mathbf{R}}^{j} - \tilde{\mathbf{R}}^{j\prime} = \pm (\tilde{\mathbf{R}}_{\alpha}^{i} - \tilde{\mathbf{R}}_{\alpha}^{i+1}) \\ 0 \qquad \text{otherwise,} \end{cases}
\end{equation}
such that, for an internal vertex, $\sum_{j=1}^{N_v} c^{ij}_\alpha \tilde{\mathbf{R}}^j=\tilde{\mathbf{R}}_\alpha+\tilde{\mathbf{R}}_\alpha^i$.
For trijunctions, there will exist $\alpha, \alpha^{\prime}, \alpha^{\prime \prime}$ for respective $i, i^{\prime}, i^{\prime \prime}$ such that $c_{\alpha}^{ij} = c_{\alpha^{ \prime}}^{i^{\prime}j} = c_{\alpha^{\prime \prime}}^{i^{\prime \prime}j}=1$, for a given $j$.  A visual representation of this geometric arrangement is given in Figure \ref{fig:geometry}.

\subsection{3.2 Cellular forces and energies}

{\color{black}We adopt} a well-established {\color{black}and widely used} vertex-based {\color{black}constitutive} model \citep{bi2015, Farhadifar:2007, fletcher2014vertex, honda1980, mao2013, nagai2001dynamic}.  We consider a monolayer of cells with identical physical properties but differing in general in size and shape.  Every cell is assumed to have a mechanical energy, $\tilde{U}_{\alpha}$, defined by
\begin{equation}
    \label{eq:single_cell_energy}
 \tilde{U}_{\alpha} = \tfrac{1}{2}\tilde{K}\left( \tilde{A}_{\alpha} - \tilde{A}_{0} \right)^{2} + \tfrac{1}{2}\tilde\Gamma \tilde{L}_{\alpha}^{2} + \tfrac{1}{2}\tilde{\Lambda} \tilde{L}_{\alpha}.
\end{equation}
The first term in \eqref{eq:single_cell_energy} models the cell's bulk compressibility, in terms of a preferred area $\tilde{A}_{0}$ and a stiffness $\tilde{K}$. The remaining terms represent the contractility of the cell periphery, via cortical actomyosin bundles and cell-to-cell adhesion.  The parameter $\tilde{\Gamma}$ represents the contractile strength while $\tilde{\Lambda}$ tunes the effective preferred cell perimeter $\tilde{L}_0=-\tilde{\Lambda}/2\tilde{\Gamma}$, such that the energy associated with the peripheral forces is of the form $\frac{1}{2}\tilde{\Gamma}(\tilde{L}-\tilde{L}_0)^2$.  The quadratic contributions to the energy as a function of perimeter and area could in principle be extended with higher-order nonlinearities. 
At the tissue level, the system is assumed to evolve down gradients of the bulk energy $\sum_{\alpha=1}^{N_c} \tilde U_{\alpha}$ from an initial disordered state.  We model the deterministic evolution by assigning a drag force (relative to the substrate on which the monolayer sits), to each vertex of cell $\alpha$, of the form $-\eta (\tilde{A}/Z_{\alpha}) \mathrm{d}\tilde{\mathbf{R}}_j/\mathrm{d}{\tilde t}$.  The drag magnitude is chosen to scale with the cell's area rather than its number of vertices {\color{black}(a natural assumption if the drag arises from physical interactions distributed across the base of the cell)} and viscous resistance to internal shear or extension is neglected.  For the time being we do not consider topological rearrangements of the network of cell edges, but return to this when discussing simulations in Section~\ref{sec:simulations}.

We nondimensionalise by scaling lengths on $\sqrt{\tilde{A}_0}$, using
\begin{equation}
\textstyle{    \tilde{A}_{\alpha} = \tilde{A}_0 A_{\alpha} \quad {\color{black}(\tilde{L}_{\alpha}, \tilde{l}^{i}_{\alpha}, \tilde{\mathbf{R}}_\alpha,\dots) = \sqrt{\tilde{A}_0} (L_{\alpha}, l^{i}_{\alpha}, {\mathbf{R}}_\alpha,\dots ),}\quad 
\tilde{U} = \tilde{K}\tilde{A}_0^2 U}, \quad \tilde{t}=\eta t/(\tilde{K} \sqrt{\tilde{A}_0}).
\end{equation}
Thus \eqref{eq:single_cell_energy} becomes
$U_{\alpha} = \tfrac{1}{2}\left( A_{\alpha} - 1 \right)^{2} + \tfrac{1}{2}\Gamma L_{\alpha}^{2} + \tfrac{1}{2} \Lambda L_{\alpha}$,
in terms of the nondimensional parameters
\begin{equation}
   \Gamma= \frac{\tilde{\Gamma}}{\tilde{K} \tilde{A}_0}, \qquad  \Lambda=\frac{\tilde{\Lambda}}{\tilde{K} \tilde{A}_0^{3/2}},
\end{equation}
where $L_{0} = -{\Lambda}/{2\Gamma}$ is the dimensionless preferred perimeter.  The total energy, $U$, of the monolayer may now be written as the sum
\begin{equation}
    \label{eq:final_nondim}
    U( \left\{ \mathbf{R}^{i}_{\alpha} \right\} ; \Gamma, \Lambda ) = \sum\limits_{\alpha=1}^{N_{c}} \left\{ \tfrac{1}{2}\left( A_{\alpha} - 1 \right)^{2} + \tfrac{1}{2} \Gamma( L_{\alpha} - L_{0} )^{2} - U_{0} \right\}
\end{equation}
where $U_{0} = {\Lambda^{2}}/{4\Gamma^{2}}$ is a constant that may be discarded as the dynamics are driven by energy gradients.
{\color{black}For later reference we define an associated pressure and tension for each cell as
\begin{equation}
\label{eq:pressten}
P_\alpha \equiv A_\alpha -1 \quad\mathrm{and}\quad T_\alpha \equiv \Gamma(L_\alpha - L_0).
\end{equation}
}

Cellular forces can be computed directly from the mechanical energy, using the fact that $\delta^{i}U_{\alpha} = \nabla^{i}U_{\alpha}\cdot\delta \mathbf{R}_{\alpha}^{i}$. The first variation of the energy with respect to the position of vertex $i$ is given by
\begin{equation}
	\delta^{i}\left\{\tfrac{1}{2}(A_{\alpha}-1)^2+\tfrac{1}{2}\Gamma(L_{\alpha}-L_0)^2\right\}=  -\mathbf{f}^{i}_{\alpha} \cdot \delta \mathbf{R}_{\alpha}^i. 
\end{equation}
$-\mathbf{f}^{i}_{\alpha} \equiv \nabla^{i} U_{\alpha}$ can be interpreted as the force required to shift vertex $i$ through $\delta \mathbf{R}_{\alpha}^i$ to do work $\delta^i U$; equivalently, $\mathbf{f}_\alpha^i$ represents the restoring force exerted at vertex $i$ by cell $\alpha$. This force can be calculated explicitly by differentiating the mechanical energy term by term. Considering first the area contribution we find
\begin{equation}
    \begin{split}
    -\nabla_{i} \tfrac{1}{2}\left( A_{\alpha} - 1 \right)^{2} &= - \left( A_{\alpha} -1 \right) \nabla_{i} A_{\alpha} 
    = -\left( A_{\alpha} - 1 \right) \nabla_{i}  \sum_{j=0}^{Z_{\alpha}-1} \tfrac{1}{2} \hat{\mathbf{z}} \cdot (\mathbf{R}_{\alpha}^j\times\mathbf{R}_{\alpha}^{j+1}) \\
    & = -\frac{1}{2}\left( A_{\alpha} - 1 \right)  (\mathbf{R}_{\alpha}^{i+1}-\mathbf{R}_{\alpha}^{i-1})\times \hat{\mathbf{z}} 
    = - P_{\alpha} \mathbf{p}_{\alpha}^{i},
    \end{split}
\end{equation}
where $P_{\alpha}$ {\color{black}is given by (\ref{eq:pressten}a)} 
and $\mathbf{p}_{\alpha}^i\equiv  \tfrac{1}{2}\left(\mathbf{n}_{\alpha}^i+\mathbf{n}_{\alpha}^{i-1}\right) = \tfrac{1}{2}(\mathbf{R}_{\alpha}^{i+1}-\mathbf{R}_{\alpha}^{i-1})\times \hat{\mathbf{z}}$ gives the direction of the bulk compressive force at node $i$.  The perimeter term gives
\begin{subequations}
    \begin{align}
    -\nabla_{i} \tfrac{1}{2} \Gamma( L_{\alpha} - L_{0} )^{2} &= -\Gamma ( L_{\alpha} - L_{0} ) \nabla_{i} L_{\alpha} 
    = -\Gamma ( L_{\alpha} - L_{0} ) \nabla_{i} \sum_{j=0}^{Z_{\alpha}-1} (\mathbf{t}_{\alpha}^j\cdot \mathbf{t}_{\alpha}^j)^{1/2} \\
    &= \Gamma ( L_{\alpha} - L_{0} ) (\hat{\mathbf{t}}_{\alpha}^{ i } - \hat{\mathbf{t}}_{\alpha}^{i-1})     = T_{\alpha} \mathbf{q}_{\alpha}^{i},
    \end{align}
\end{subequations}
where $T_{\alpha}$ {\color{black}(see (\ref{eq:pressten}b))} 
represents a tension and $\mathbf{q}_{\alpha}^{i} \equiv \hat{\mathbf{t}}_{\alpha}^{ i } - \hat{\mathbf{t}}_{\alpha}^{i-1}$ represents the direction of the inward force due to stretching of the cell perimeter. Thus the force at vertex $i$ can be written
\begin{equation}
    \mathbf{f}_{\alpha}^{i} = - P_{\alpha} \mathbf{p}_{\alpha}^{i} + T_{\alpha} \mathbf{q}_{\alpha}^{i}.
\end{equation}
{\color{black}The analogous force for a vertex model lacking the $\tilde{L}^2_\alpha$ term in (\ref{eq:single_cell_energy}) is given in \cite{spencer2017}.}

$\mathbf{f}_{\alpha}^{i}$ represents the force generated when perturbing the vertex of a cell in isolation. For the case of a monolayer, each vertex will have a contribution from the three cells attached to it (or fewer, if the cell is at the periphery of the monolayer). Thus the net force on vertex $j$, $\mathbf{f}^{j}$, will be given by the sum of the contributions from each cell attached to it as
\begin{equation}
    \mathbf{f}^{j} = \sum\limits_{\alpha=1}^{N_{c}} \sum_{i=0}^{Z_{\alpha}-1}  (- P_{\alpha} \mathbf{p}_{\alpha}^{i} + T_{\alpha} \mathbf{q}_{\alpha}^{i})c_{\alpha}^{ij},
\end{equation}
where $c_{\alpha}^{ij}$ ensures that, although the summation is over all cells, we count only the contributions from the cells connected to vertex $j$. More specifically, if cells $\alpha$, $\alpha'$ and $\alpha''$ meet at junction $j$, with anticlockwise tangents $\mathbf{t}$, $\mathbf{t}'$, $\mathbf{t}''$ emerging from the vertex with normals (pointing clockwise) $\mathbf{n}$, $\mathbf{n}'$, $\mathbf{n}''$ orthogonal to each tangent, the net force at the vertex can be written
\begin{multline}
\mathbf{f}^j=\mathbf{t}(T_{\alpha}+T_{\alpha''})+\mathbf{t}'(T_{\alpha'}+T_{\alpha}) +\mathbf{t}''(T_{\alpha''}+T_{\alpha'}) \\
+\tfrac{1}{2}\left[\mathbf{n}(P_{\alpha}-P_{\alpha''})+\mathbf{n}'(P_{\alpha'}-P_{\alpha})+\mathbf{n}''(P_{\alpha''}-P_{\alpha'})\right].
\end{multline}
The tangential forces show how each edge is a composite structure with tension contributions from two adjacent cells.  The factor of $\tfrac{1}{2}$ in the pressure terms reflects the fact that the force due to pressure acting on any edge is distributed equally between each vertex bounding the edge.  The tensions and pressures depend on the total area and perimeter of each neighbouring cell {\color{black}via (\ref{eq:pressten})}.   For vertices at the periphery of the monolayer, bordering cells $\alpha$ and $\alpha'$, we write $P_{\alpha''}=P_{\mathrm{ext}}$ (an imposed isotropic stress) and set $T_{\alpha''}=0$, so that
\begin{equation}
\label{eq:periph}
\mathbf{f}^j=\mathbf{t} T_{\alpha}+\mathbf{t}'(T_{\alpha'}+T_{\alpha}) +\mathbf{t}'' T_{\alpha'} 
+\tfrac{1}{2}\left[P_{\alpha}\mathbf{n}+\mathbf{n}'(P_{\alpha'}-P_{\alpha})-P_{\alpha'}\mathbf{n}''\right]
+\tfrac{1}{2}P_{\mathrm{ext}}(\mathbf{n}''-\mathbf{n}).
\end{equation}
We use this relationship below when considering the boundary conditions at the edge of a monolayer.

When the system is out of equilibrium, the net force at each internal vertex is 
\begin{equation}
\mathbf{F}^j=\mathbf{f}^j-\left(\frac{A_\alpha}{Z_\alpha}+\frac{A_{\alpha'}}{Z_{\alpha'}}+\frac{A_{\alpha''}}{Z_{\alpha''}}\right) \dot{\mathbf{R}}^j,
\end{equation}
where the term proportional to $\dot{\mathbf{R}}_j$ is a viscous drag having 
contributions from the three cells at the trijunction; the dot denotes a time derivative.  Writing $\dot{\mathbf{R}}^j=\dot{\mathbf{R}}_\alpha+\dot{\mathbf{R}}_\alpha^i$, the drag can be considered as representing an internal dashpot within each cell connecting the cell centre to the vertex plus a drag on each cell centre. 
Thus the net force on cell $\alpha$ becomes
\begin{equation}
\mathbf{F}_\alpha=-\sum_{i=0}^{Z_{\alpha}-1} \left(\mathbf{f}_\alpha^i -(A_\alpha/Z_\alpha) (\dot{\mathbf{R}}_\alpha+\dot{\mathbf{R}}_\alpha^i) \right)
=-\mathbf{f}_\alpha+A_\alpha \dot{\mathbf{R}}_\alpha
\label{eq:cellforce}
\end{equation}
where $\mathbf{f}_\alpha=\sum_{i=0}^{Z_{\alpha}-1} \mathbf{f}_{\alpha}^i$, noting that $\sum_{i=0}^{Z_\alpha-1} \dot{\mathbf{R}}_{\alpha}^i=\mathbf{0}$.  Since inertia is negligible, the net force on any vertex and on any cell must vanish, $\mathbf{F}^j=\mathbf{0}$ and $\mathbf{F}_\alpha=\mathbf{0}$.  The former condition defines the $N_v$ coupled evolution equations of the network vertices.  When the system is in equilibrium, this simplifies to $\mathbf{f}^j=\mathbf{0}$, $\textbf{f}_{\alpha}=\textbf{0}$.  Likewise the net torque on cell $\alpha$,
\begin{equation}
\mathbf{T}_\alpha=-\sum_{i=0}^{Z_{\alpha}-1} \mathbf{R}_\alpha^i\times \left(\mathbf{f}_\alpha^i -(A_\alpha/Z_\alpha) \left(\dot{\mathbf{R}}_\alpha+\dot{\mathbf{R}}_\alpha^i\right)\right)
=-\sum_{i=0}^{Z_{\alpha}-1} \mathbf{R}_\alpha^i\times \left(\mathbf{f}_\alpha^i -(A_\alpha/Z_\alpha)\dot{\mathbf{R}}_\alpha^i \right),
\label{eq:celltorque}
\end{equation}
must satisfy $\mathbf{T}_\alpha=\mathbf{0}$.

\subsection{3.3 The stress tensor of a cell}

For a tensor $\boldsymbol{\sigma}$ that is symmetric and divergence-free, defined over an area $\mathcal{A}$ with perimeter $\mathcal{S}$, we have $\boldsymbol{\sigma}=\nabla\cdot (\mathbf{R}\otimes\boldsymbol{\sigma})$, {\color{black}where $\mathbf{R}$ is an arbitrary position vector}. Thus taking an area integral and applying the divergence theorem gives \citep{norris2014}
\begin{equation}
	\label{eq:stress_tensor_origin}
	\begin{split}
	\int_{\mathcal{A}}\boldsymbol{\sigma} \, \text{d}A &= \int_{\mathcal{A}} \nabla \cdot (\mathbf{R} \otimes \boldsymbol{\sigma}) \, \text{d}A 
	= \oint_{\mathcal{S}} \mathbf{R} \otimes \boldsymbol{\sigma} \cdot \mathbf{n} \, \text{d}S.
	\end{split}
\end{equation}
We use this weak formulation to derive the stress tensor of the monolayer, taking the stress to be uniform over each cell.  The forces acting on cell $\alpha$ are distributed around the vertices, so that taking $\mathcal{A}=\mathcal{A}_\alpha$ (the domain of cell $\alpha$), (\ref{eq:stress_tensor_origin}) motivates the definition of the cell stress $\boldsymbol{\sigma}_\alpha$ as 
\begin{subequations}
\label{eq:cellstrdef}
\begin{align}
A_{\alpha} \boldsymbol{\sigma}_{\alpha} &= \sum_{j=1}^{N_v} \sum_{i=0}^{Z_{\alpha}-1} c_{\alpha}^{ij} \mathbf{R}^j\otimes \mathbf{F}_{\alpha}^i =\sum_{i=0}^{Z_{\alpha}-1} (\mathbf{R}_\alpha+\mathbf{R}_\alpha^i)\otimes \mathbf{F}_{\alpha}^i \\
 &=\sum_{i=0}^{Z_{\alpha}-1} \mathbf{R}_\alpha^i\otimes (\mathbf{f}_{\alpha}^i -(A_\alpha/Z_\alpha) (\dot{\mathbf{R}}_\alpha+\dot{\mathbf{R}}_\alpha^i))
 =\sum_{i=0}^{Z_{\alpha}-1} \mathbf{R}_\alpha^i\otimes \mathbf{f}_{\alpha}^i -(A_\alpha/Z_\alpha) \sum_{i=0}^{Z_{\alpha}-1} \mathbf{R}_\alpha^i\otimes \dot{\mathbf{R}}_\alpha^i.
\end{align}
\end{subequations}
This reveals conservative (elastic) and dissipative (viscous) contributions to the stress.  The former is
 \begin{equation}
    \label{eq:stress_def}
  \sum_{i=0}^{Z_{\alpha}-1} \mathbf{R}_{\alpha}^i\otimes \mathbf{f}_{\alpha}^i 
    = \sum_{i=0}^{Z_{\alpha}-1} \mathbf{R}_{\alpha}^i\otimes (-P_{\alpha}\mathbf{p}_{\alpha}^i + T_{\alpha}\mathbf{q}_{\alpha}^i).
\end{equation}
If the cell is in equilibrium and under zero net torque, then $\sum_{i=0}^{Z_\alpha -1} \mathbf{R}_\alpha^i \times \mathbf{f}_{\alpha}^i=\textbf{0}$ (see \eqref{eq:celltorque}), ensuring that this contribution to $\boldsymbol{\sigma}_\alpha$ is symmetric; the symmetry of (\ref{eq:stress_def}) is confirmed below.   {\color{black}Likewise the absence of torque on a cell due to drag in (\ref{eq:celltorque}) requires} the dissipative component of the stress to be symmetric, {\color{black}allowing us to redefine} the final term in (\ref{eq:cellstrdef}b) as
\begin{equation}
-\frac{1}{2Z_\alpha}  \sum_{i=0}^{Z_{\alpha}-1} (\mathbf{R}_\alpha^i\otimes \dot{\mathbf{R}}_\alpha^i+ \dot{\mathbf{R}}_\alpha^i\otimes {\mathbf{R}}_\alpha^i)\equiv-\tfrac{1}{2}\dot{\mathsf{S}}_\alpha,
\end{equation}
{\color{black}where $\mathsf{S}_\alpha$ is the dimensionless shape tensor based on vertex location.}

We simplify (\ref{eq:stress_def}) by making use of two geometric identities, established in Appendix A, namely 
\begin{equation}
\label{eq:geomid}
   \sum_{i=0}^{Z_{\alpha}-1} \mathbf{R}_{\alpha}^i\otimes \mathbf{p}_{\alpha}^i 
    =  A_{\alpha} \mathsf{I},  \qquad
    \sum_{i=0}^{Z_{\alpha}-1} \mathbf{R}_{\alpha}^i\otimes \mathbf{q}_{\alpha}^i 
    	= - \sum_{i=0}^{Z_{\alpha}-1} \hat{\mathbf{t}}_{\alpha}^{i} \otimes \mathbf{t}_{\alpha}^{i},
\end{equation}
both of which are symmetric (recall $\mathbf{t}_\alpha^i=l_\alpha^{i}\hat{\mathbf{t}}_\alpha^i$).  Noting that $\mathrm{Tr}(\sum\nolimits_{i=0}^{Z_{\alpha}-1} \hat{\mathbf{t}}_{\alpha}^i \otimes \mathbf{t}_{\alpha}^i) =\sum\nolimits_{i=0}^{Z_{\alpha}-1} \hat{\mathbf{t}}_{\alpha}^i \cdot \mathbf{t}_{\alpha}^i =\sum\nolimits_{i=0}^{Z_{\alpha}-1} l_\alpha^i=L_\alpha$, we can then express the stress of cell $\alpha$ as 
\begin{equation}
    \label{eq:full_stress_inc_dissipative}
    \boldsymbol{\sigma}_{\alpha} = -P^{\text{eff}}_{\alpha}\mathsf{I} + T_{\alpha} {\mathsf{J}}_{\alpha} - \tfrac{1}{2}\dot{\mathsf{S}}_\alpha. 
\end{equation}
Here the elastic components of the stress have been written in terms of an isotropic and deviatoric component.  The former defines the effective cell pressure, which has contributions from the cell's bulk and the perimeter (in Young--Laplace form, {\color{black}with an effective radius of curvature $2A_\alpha/L_\alpha$}) as
\begin{equation}
P^{\text{eff}}_{\alpha} = P_{\alpha} + \frac{T_{\alpha} L_{\alpha}}{2 A_{\alpha}}.
\label{eq:peff}
\end{equation}
We will see below how the competition between bulk pressure and cortical forces can stiffen the monolayer.  The traceless contribution to the cell stress is
\begin{equation}
    \label{eq:traceless_stress}
    \begin{split}
    {\mathsf{J}}_{\alpha} =  
    \frac{1}{A_{\alpha}} \left( \tfrac{1}{2} L_{\alpha}\mathsf{I} - \sum\limits_{i=0}^{Z_{\alpha}-1} l^{i}_{\alpha}  \hat{\text{\textbf{t}}}_{\alpha}^{i} \otimes \hat{\text{\textbf{t}}}_{\alpha}^{i}   \right).
    \end{split}
\end{equation}

\subsection{3.4 Relating cell stress and shape}

We can now explore the relationship between the principal axes of cell shape and stress by considering the commutativity of $\boldsymbol{\sigma}_{\alpha}$ and $\mathsf{S}_{\alpha}$. The tensors will share an eigenbasis, implying that their principal axes align, if and only if they commute. Having separated the stress tensor (\ref{eq:full_stress_inc_dissipative}) into an isotropic and deviatoric component however, we require only that $\mathsf{S}_{\alpha}\mathsf{J}_{\alpha} = \mathsf{J}_{\alpha}\mathsf{S}_{\alpha}$ {\color{black}and $\mathsf{S}_{\alpha}\dot{\mathsf{S}}_{\alpha} = \dot{\mathsf{S}}_{\alpha}\mathsf{S}_{\alpha}$}, which is established {\color{black}via direct algebraic manipulation} in Appendix~\ref{sec:s_j_align}.   Figure \ref{fig:stress_shape_align} provides a computational illustration of this {\color{black}mathematical} result {\color{black} for a disordered monolayer in equilibrium}; details of the simulation scheme are are given in Section~\ref{sec:simulations}.  Thus, for an individual cell, the principal axes of stress and shape align {\color{black}(both quantities being defined directly in terms of cell vertex locations).}  Equivalently, within the present model, cells that are elongated experience a {\color{black}local} stress field that is oriented exactly with the direction of elongation.  The consequences of this observation are discussed below.

\begin{figure}
	\centering
	\includegraphics[width=0.5\textwidth]{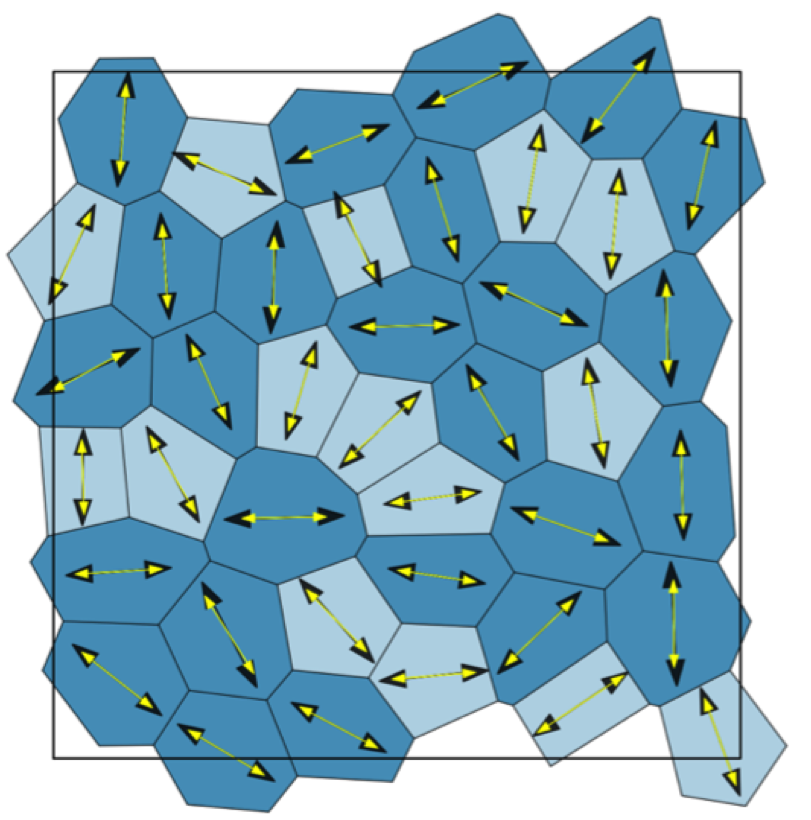}
	\caption{Computational validation of the predicted alignment between principal axis of stress and shape, for $(\Lambda, \Gamma) = (-0.2, 0.1)$. The initial cell array was were generated using a Voronoi tessellation and then relaxed to equilibrium using periodic boundary conditions. The eigenvectors corresponding the the principal eigenvalue of $\boldsymbol{\sigma}_{\alpha}$ and $\mathsf{S}$ are plotted in black and yellow respectively. Darker cells have $P_{\alpha}^{\text{eff}} > 0$ (net tension); lighter cells have $P_{\alpha}^{\text{eff}} < 0$ (net compression).}
    \label{fig:stress_shape_align}
\end{figure}

\subsection{3.5 Stress of the monolayer}

We now return to (\ref{eq:stress_tensor_origin}), taking the domain $\mathcal{A}$ in (\ref{eq:stress_tensor_origin}) to cover multiple cells.  The area integral can be evaluated over each cell to give a formulation for the `tissue' stress over a simply connected region $R$ of the monolayer $\boldsymbol{\sigma}^R$ as 
\begin{equation}
\label{eq:tissuestr}
\left (\sum_\alpha A_\alpha\right ) \boldsymbol{\sigma}^R=\sum_\alpha \boldsymbol{\sigma}_\alpha A_\alpha,   
\end{equation}
summing over cells in $R$.   The components of the first two terms on the right-hand side of (\ref{eq:full_stress_inc_dissipative}) that are proportional to $T_\alpha$ at the cell level, and their area-weighted sum in (\ref{eq:tissuestr}), are analogous to an expression derived by \cite{batchelor1970} for a suspension of particles having interfacial tension.  Equivalent expressions for the equilibrium stress of the present model based on Batchelor's formulation have been given by  \cite{ishihara2012} and \cite{guirao2015}.  

For now let us take $R$ to be the whole monolayer.  The line integral in (\ref{eq:stress_tensor_origin}) can be evaluated by setting
\begin{equation}
\oint_{\mathcal{P}} \mathbf{R}\otimes \boldsymbol{\sigma} \cdot \mathbf{n}\,\mathrm{d}S= \sum_\alpha \oint_{\partial\mathcal{A}_\alpha} \mathbf{R}\otimes \boldsymbol{\sigma} \cdot \mathbf{n}\,\mathrm{d}S
\end{equation}
since $\mathbf{F}^j=\mathbf{0}$ at all internal vertices.  Let $k=0,1,\dots,N_p-1$ label the peripheral vertices, let peripheral normals $\mathbf{n}_{k-1}$ and $\mathbf{n}_k$ border vertex $k$ and let $\mathbf{p}_k=\tfrac{1}{2}(\mathbf{n}_{k-1}+\mathbf{n}_k)$.  Since the periphery is a closed curve, its sum of tangents vanish, hence its sum of normals vanish, hence $\sum_{k=0}^{N_p-1} \mathbf{p}_k=\mathbf{0}$.  Let $\mathbf{R}_0$ be the centroid of the monolayer, and write $\mathbf{R}^k=\mathbf{R}_0+\mathbf{R}_0^k$, so that $\sum_{k=0}^{N_p-1} \mathbf{R}_0^k=\mathbf{0}$.  Assuming the pressure is $P_{\mathrm{ext}}$ {\color{black}uniformly} around the periphery, the force balance at the peripheral vertices (\ref{eq:periph}) gives
\begin{equation}
-\tfrac{1}{2}P_{\mathrm{ext}}\sum_{k=0}^{N_p-1} \mathbf{R}^k\otimes   \left(\mathbf{n}^k+\mathbf{n}^{k-1}\right)
=-P_{\mathrm{ext}} \sum_{k=0}^{N_p-1} \mathbf{R}^k\otimes   \mathbf{p}^k
=-P_{\mathrm{ext}} \sum_{k=0}^{N_p-1} \mathbf{R}_0^k\otimes   \mathbf{p}^k=-P_{\mathrm{ext}}A\mathsf{I}
\end{equation}
where the final expression results from (\ref{eq:geomid}a) and $A=\sum_{\alpha=1}^{N_c}A_{\alpha}$. Thus $\sum_\alpha \boldsymbol{\sigma}_\alpha A_\alpha=-P_{\mathrm{ext}}A\mathsf{I}$, \hbox{i.e.}
\begin{equation}
\sum_{\alpha=1}^{N_c} A_\alpha \left(-P^{\mathrm{eff}}_\alpha \mathsf{I} + T_\alpha {\mathsf{J}}_\alpha - \tfrac{1}{2}\dot{\mathsf{S}}_\alpha\right)=-P_{\mathrm{ext}}A\mathsf{I}.
\end{equation}
Taking the trace of this sum gives
\begin{equation}
\sum_{\alpha=1}^{N_c} A_\alpha P^{\mathrm{eff}}_\alpha = A P_{\mathrm{ext}}- \sum_{\alpha=1}^{N_c}  \frac{A_\alpha}{4}\mathrm{Tr}(\dot{\mathsf{S}}_\alpha),
\end{equation}
which describes the relaxation of the area of the monolayer to its equilibrium.  Once in equilibrium, the system must satisfy
\begin{equation}
\sum_{\alpha=1}^{N_c} A_\alpha P^{\mathrm{eff}}_\alpha =A P_{\mathrm{ext}}, \quad 
\sum_{\alpha=1}^{N_c} T_\alpha {\mathsf{J}}_\alpha=\mathsf{0}.
\label{eq:weight}
\end{equation}
A disordered distribution of cells within an equilibrium monolayer will have a range of values of $P_\alpha^{\mathrm{eff}}$, and non-isotropic cells will have deviatoric contributions to their stress, but the whole population must satisfy the weighted sums (\ref{eq:weight}).  For an isolated monolayer that is in equilibrium under zero external loading (the condition relevant to Section~\ref{sec:experiments}), we must therefore impose 
\begin{equation}
\sum_{\alpha=1}^{N_c} A_\alpha P^{\mathrm{eff}}_\alpha ={0}.
\label{eq:zeroload}
\end{equation}

\subsection{3.6 Elastic moduli}

When the cells are identical hexagons, the stress at the tissue level under the present model (neglecting friction) is equivalent to that of linear elasticity when considering small perturbations about the unstressed state \citep{Murisic:2015}.  We can therefore use the expressions for stress at cell (\ref{eq:full_stress_inc_dissipative}) and tissue (\ref{eq:tissuestr}) level to recover expressions for the associated elastic moduli. 

Taking $P_{\mathrm{ext}}=0$ in a base state, imposing (\ref{eq:zeroload}), we consider an isotropic expansion of a {\color{black}disordered} monolayer of magnitude $1+\epsilon$ where $\epsilon \ll 1$, so that $L_\alpha$ maps to $(1+\epsilon) L_\alpha$, $A_\alpha$ maps to $(1+2\epsilon)A_\alpha$ and so on.  Linearising about the base state, the dimensional bulk modulus, $\tilde{K}\tilde{A}_0 K$, of the monolayer is given by 
\begin{equation}
K=A\frac{\mathrm{d}P_{\mathrm{ext}}}{\mathrm{d}A}\Bigg\vert_{P_{\mathrm{ext}}=0}=\sum_{\alpha=1}^{N_c} \frac{A_\alpha}{2A} \left[ 2A_{\alpha}+\frac{\Gamma L_0 L_\alpha}{2A_\alpha} \right],
\label{eq:bulk}
\end{equation}
using (\ref{eq:peff}).   This prediction holds for a disordered network of cells, {\color{black}and therefore provides a direct means of determining the variability of bulk modulus over different realisations of the monolayer}.  When simplified to the special case of a hexagonal monolayer, for which $L_\alpha/\sqrt{A_\alpha}=\mu_{6} \equiv 2\sqrt{2\sqrt{3}}\approx 3.72$ for all $\alpha$, \eqref{eq:bulk} reduces in dimensionless form to
\begin{equation}
	\label{eq:bulk_hex}
	K = A_\alpha - \frac{\Lambda \mu_{6}}{8 \sqrt{A_\alpha}},
\end{equation}
in agreement with \cite{Murisic:2015} and \cite{Staple:2010}.  $K$ remains positive for $\Lambda<0$, but can become zero at $\Lambda=(8/\mu_6)^{2/3}$ when $A=1$.  The dimensional shear modulus, $\tilde{K}\tilde{A}_0G$, for the special case of a monolayer {\color{black}of identical hexagonal cells} is shown in Appendix~\ref{sec:shear} to be given by
\begin{equation}
	\label{eq:shear_hex}
    G = 3 \sqrt{3} \Gamma\left( 1 - \frac{L_{0}}{L_\alpha} \right),
\end{equation}
which is also equivalent to the shear modulus derived by \cite{Murisic:2015} (but differs, as they showed, with \cite{Staple:2010}).  Equation (\ref{eq:shear_hex}) illustrates how $L$ must exceed $L_0$, \hbox{i.e.} cell walls must be under tension, in order for the monolayer to resist shear.  {\color{black}Prediction of the shear modulus for the disordered monolayer is much less straightforward; estimates (for a disordered dry foam) are reviewed in \cite{Kruyt2007}.}

\subsection{3.7 Mapping parameter space}
\label{sec:param_space}

\begin{figure}
	\centering
	\includegraphics[width=0.99\textwidth]{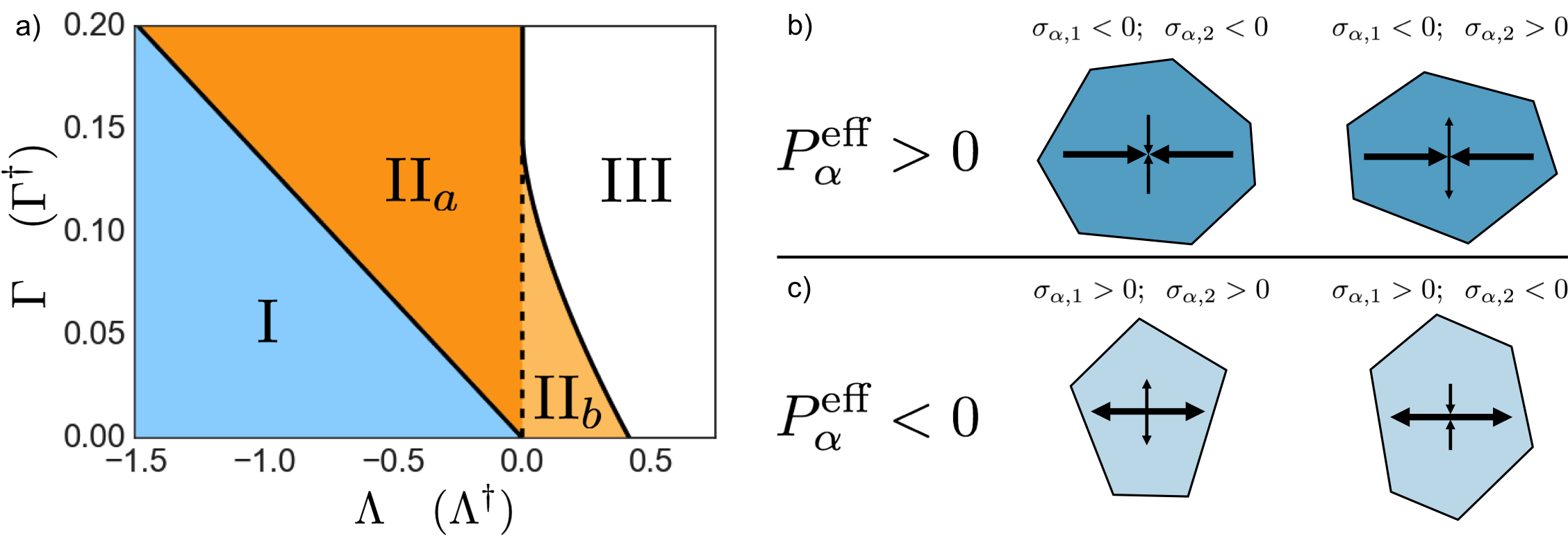}
		\caption{(a) $(\Lambda, \Gamma)$-parameter space, showing boundaries for a uniform hexagonal array (following \cite{Farhadifar:2007}). Region I represents a 'soft' network with no shear resistance, bounded by (\ref{eq:1}); $P^{\text{eff}}=0$ has a single positive root in region II$_a$ and two positive roots in region $\text{II}_{b}$.  The network collapses in Region III, which is bounded by $\Lambda=0$ and (\ref{eq:2}). {\color{black}The transformation (\ref{eq:trans}) allows $(\Lambda,\Gamma)$ to be replaced by $(\Lambda^\dag,\Gamma^\dag)$ in order to describe cases for which $P_{\mathrm{ext}}\neq 0$.} (b,c) Classification of cell stress configurations {\color{black}in a disordered monolayer}, showing representative cell shapes.  Larger (smaller) arrows indicate the orientation of the eigenvector associated with the eigenvalue of the stress tensor having larger (smaller) magnitude, where $\vert\sigma_{\alpha,1}\vert\geq \vert\sigma _{\alpha, 2}\vert \geq 0$.  Inward- (outward)-pointing arrows indicate the tension (compression) generated by the cell.}
    \label{fig:ground_state}
\end{figure}

Prior to presenting simulations, it is helpful to review the main features of parameter space \citep{Farhadifar:2007, Staple:2010}.   Recall from (\ref{eq:peff}) that $P^{\mathrm{eff}}_\alpha=P^{\mathrm{eff}}(A_{\alpha},L_{\alpha})$ where 
\begin{equation}
P^{\mathrm{eff}}(A,L)=A-1+\Gamma (L-L_0)L/2A.
\end{equation}
For a perfect N-gon, with perimeter and area satisfying $L=\mu_N \sqrt{A}$ where $\mu_N=2(N \tan(\pi/N))^{1/2}$, 
\begin{equation}
	\label{eq:peff_hex}
    P^{\text{eff}}_N(A) = A-1+\tfrac{1}{2}\Gamma\left(\mu_N^2 -\frac{L_0 \mu_N}{\sqrt{A}} \right)\equiv A - 1 + \frac{\Gamma\mu_N^{2}}{2} + \frac{\Lambda\mu_N}{4\sqrt{A}}.
\end{equation}
We define $A_N^*(\Gamma$, $\Lambda)$ to satisfy $P^{\text{eff}}_N(A_N^*)=0$, {\color{black}to satisfy the constraint (\ref{eq:zeroload})}.  Thus for hexagons, for example, $A_6^*=1$ when $L_0=\mu_6$, \hbox{i.e.}
\begin{equation}
\Lambda=-2\mu_6 \Gamma, \quad \Gamma<0.
\label{eq:1}
\end{equation}
Analysis of the cubic $\sqrt{A}P^{\text{eff}}_6$ as a function of $\sqrt{A}$ reveals that it is monotonic (implying a single root of $P^{\text{eff}}_6=0$) for $\Gamma>2/\mu_6^2$; a positive root exists provided $\Lambda<0$ for $\Gamma>0$ that satisfies $A_6^*=1$ along (\ref{eq:1}).
For $\Lambda>0$, the cubic has repeated roots along
\begin{equation}
\Lambda=\frac{8}{3^{3/2}\mu_6}\left(1-\tfrac{1}{2}\Gamma \mu_6^2\right)^{3/2}, \quad 0<\Gamma<2/\mu_6^2.
\label{eq:2}
\end{equation}
As a consequence the parameter map shown in Figure~\ref{fig:ground_state} can be drawn \citep{Farhadifar:2007}, with the boundary between regions I and $\text{II}_a$ defined by (\ref{eq:1}), that between regions $\text{II}_a$ and III by $\Lambda=0$ and $\Gamma>2/\mu_6^2$ and that between regions $\text{II}_{b}$ and III by \eqref{eq:2}.   We will focus attention below on region II, in which at least one stress-free equilibrium state exists (for hexagons) with positive shear modulus.  Along the region I/$\text{II}_a$ boundary, hexagons have $P=0$ ($A=1$) and $T=0$ ($L_0=L=\mu_6$) and the monolayer loses any resistance to shear (from \eqref{eq:shear_hex}).  (In a disordered monolayer, the rigidity transition to a floppy region-I state has been shown to arise closer to $L_0=\mu_5\approx 3.81$ \citep{bi2015}.)  Approaching the region $\text{II}_a$/III boundary, the equilibrium cell area approaches $A=0$; two possible equilibria exist in region IIb, coalescing at positive $A$ along the region $\text{II}_b$/III boundary.

{\color{black} For later reference, we note that for a periodic array of hexagons under an external load $P_{\mathrm{ext}}$ (for which $P^{\mathrm{eff}}_\alpha=P_{\mathrm{ext}}$ in (\ref{eq:weight})), we may define (for $P_{\mathrm{ext}}>-1$)
\begin{equation}
    A^{\dag} = {A}/{(1+P_{\text{ext}})} \qquad \Gamma^{\dag} = {\Gamma}/{(1+P_{\text{ext}})} \qquad \Lambda^{\dag} = {\Lambda}/{(1+P_{\text{ext}})^{\tfrac{3}{2}}},
    \label{eq:trans}
\end{equation}
such that if $P_6^\mathrm{eff}(A;\Gamma,\Lambda)=P_{\mathrm{ext}}$ in (\ref{eq:peff_hex}) then $P^\mathrm{eff}_6(A^{\dag}; \Gamma^\dag, \Lambda^\dag)=0$.  This simple scaling symmetry of (\ref{eq:peff_hex}) allows the axes of Figure~\ref{fig:ground_state}(a) to be replaced with $\Lambda^\dag$ and $\Gamma^\dag$ in order to encompass externally-loaded monolayers subject to non-zero $P_{\mathrm{ext}}$.
}

Figure~\ref{fig:ground_state}(b,c) illustrates four distinct classes of equilibrium cell shape and stress that arise in simulations {\color{black}of disordered monolayers}, distinguished by the signs of the eigenvalues $(\sigma_{\alpha,1}, \sigma_{\alpha,2})$ of the cell stress tensor; recall that the corresponding eigenvectors align with the principal axes of the shape tensor $\mathsf{S}_\alpha$.  When $P^{\text{eff}}_\alpha= -\mathrm{Tr}(\boldsymbol{\sigma}_\alpha)\equiv -(\sigma_{\alpha,1}+\sigma_{\alpha,2})>0$ (represented by darker cells, Figure~\ref{fig:ground_state}(b), the cell is enlarged and under net tension: both eigenvalues of the stress tensor are negative when the cell is rounder, although one can be positive when the cell is more elongated.  Likewise when $P^{\text{eff}}_\alpha<0$ (lighter cells, Figure~\ref{fig:ground_state}(c), the cell is smaller and under net compression: both eigenvalues of the stress tensor are positive when the cell is rounder, although one can be negative when the cell is more elongated.

\subsection{3.8 Simulation methodology}

The majority of computational modelling was performed in Python, with some processes sent through C where Python struggled with performance. The cells were described as an oriented graph using the graph-tool module for Python \citep{graph-tool}. The algorithms and core data structures of graph-tool are written in C++, thus its performance in memory and computation is comparable to that of pure C++. The energy minimisation was performed using a conjugate gradient method from the {\tt{scipy}} library.  

Simulations were performed in a square box of side $\mathcal{L}$, imposing periodic boundary conditions.  A Mat\'ern type II random sampling process was used to identify $N_c$ initial cell centres within the box, giving mean cell area $\bar{A}=\mathcal{L}^2/N_c$, chosen to match $A_6^*$ (given that hexagons are the most frequently observed polygonal class in monolayers \citep{Gibson:2006}). A Voronoi tessellation was constructed between the points (and their periodic extensions) to define an initial network of edges and vertices.  The system was then relaxed towards the nearest energy minimum. 
If the length of any edge fell beneath $0.1\sqrt{A_6^*}$  (taking the larger value of $\tilde{A}^*_6$ in Region $\text{II}_b$), a T1 transition (or intercalation) was implemented and relaxation proceeded further {\color{black}(see \cite{spencer2017} for a more refined treatment of this process)}.  If the area of a 3-sided cell fell beneath $0.3A_6^*$ (again taking the larger value of $\tilde{A}^*_6$ in region $\text{II}_b$), the cell was removed via a T2 transition (extrusion).  A small isotropic expansion or contraction of the network and the bounding box was used to satisfy the zero-load condition (\ref{eq:zeroload}) within an prescribed tolerance.  The initial disorder produced a distribution of values of $P_\alpha^{\mathrm{eff}}$ across the cell population. 

\section{Results}
\label{sec:simulations}

\begin{figure}
	\centering
	\includegraphics[width=0.95\textwidth]{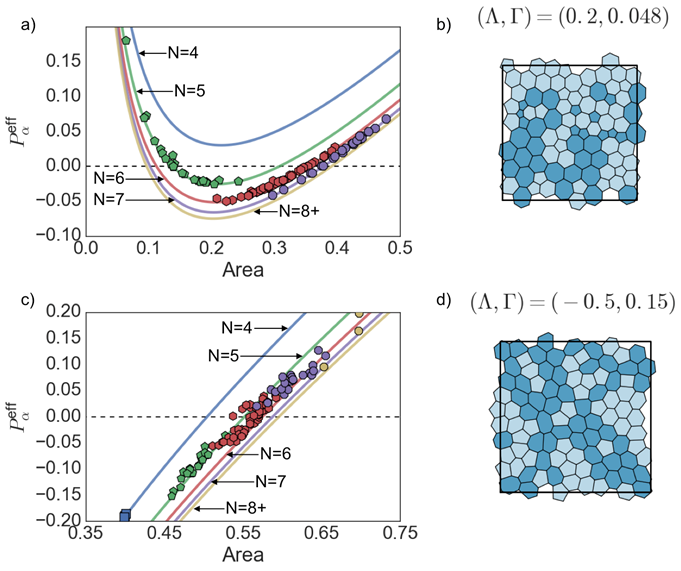} 
	\caption{(a,c) Curves show $P^{\text{eff}}_N$ defined in (\ref{eq:peff_hex}) plotted against cell area for perfect N-gons, using $(\Lambda, \Gamma) = (-0.5, 0.15)$ ($L_0=3.33$, a,b) and $(\Lambda, \Gamma) = (0.2, 0.048)$ ($L_0=-2.08$, c,d).  Symbols show $P^{\text{eff}}_\alpha$ defined in (\ref{eq:peff}) for computationally simulated cells, with shapes displayed in (b,d). Darker (lighter) cells in (b,d) have $P_{\alpha}^{\text{eff}} > 0$ ($<0$).}
    \label{fig:non_monotonic_peff}
\end{figure}

Simulations for $\Lambda>0$ and $\Lambda<0$  are illustrated in Figure \ref{fig:non_monotonic_peff}(a,b) and (c,d) respectively.  In both examples, the $P^{\mathrm{eff}}_\alpha$ for individual cells in the disordered monolayer lie close to $P^{\text{eff}}_N$, the values for perfect polygons, suggesting that $P^{\mathrm{eff}}_\alpha$ can be well predicted by a cell's area and its polygonal class.  $P^{\mathrm{eff}}_N$ is monotonic in cell area when $\Lambda<0$ ($L_0>0$), whereas it has a turning point for $\Lambda>0$. Despite the potential for bistability in the latter case, cells in a disordered array lie on both branches of the $P^{\text{eff}}_N$ curves.  In both examples, the mean cell area over the monolayer lies below unity, implying that cells lie below their equilibrium area: each cell is held at this level by cortical tension, as the cell perimeters exceed the target value $L_0$.    Simulations show that pentagons are smaller on average than heptagons; when $\Lambda<0$ pentagons have $P^{\text{eff}}_\alpha<0$ and heptagons have $P^{\text{eff}}_\alpha>0$ (Figure~\ref{fig:non_monotonic_peff}c); in contrast, for $\Lambda<0$ both sets of cells cluster around $P^{\text{eff}}_\alpha=0$ (Figure~\ref{fig:non_monotonic_peff}a). 

\begin{figure}
	\centering
	\includegraphics[width=0.95\textwidth]{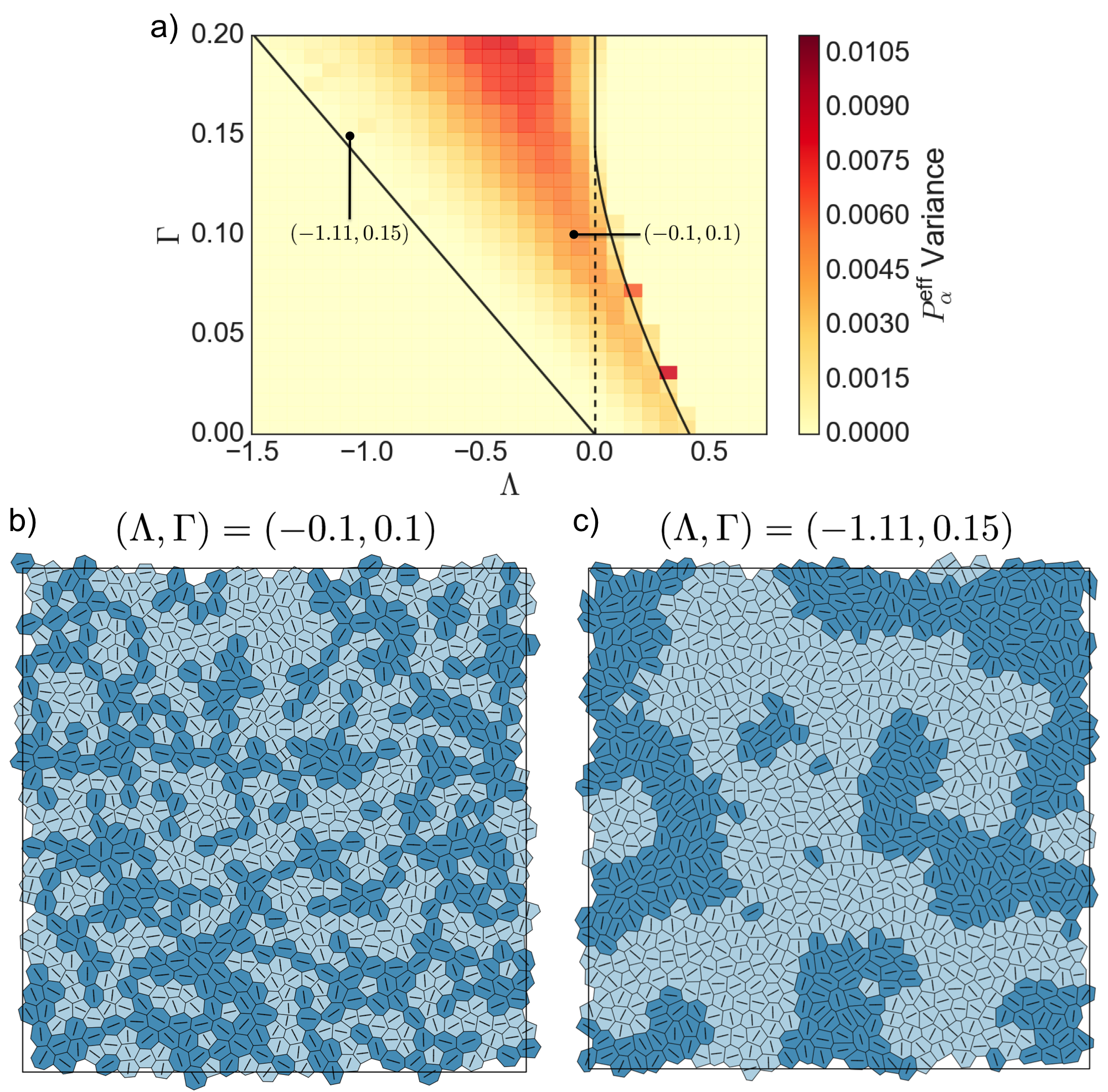} 
	\caption{(a)  A map of the variance of $P^{\text{eff}}_\alpha$ at discrete locations within region II of $(\Lambda,\Gamma)$-parameter space.  Lines show the boundaries for a hexagonal network, as in Figure~\ref{fig:ground_state}(a).  The dark squares along the region II$_b$/III boundary are artefacts, reflecting the co-existence of cells with small and large areas near this boundary.  Each datapoint is taken from {\color{black}5 realisations} of a monolayer with 800 cells.  (b) An individual monolayer realisation for $\Lambda=-0.1$, $\Gamma=0.1$, $P_{\mathrm{ext}}=0$ with 800 cells.  Darker (lighter) shading denotes cells with $P^{\text{eff}}>0$ $(<0)$. Line segments indicate the principal axis of the shape and stress tensor for each cell, coincident with the heavy arrows in Figure~\ref{fig:ground_state}(b), \hbox{i.e.} aligned with the stress eigenvector associated with the eigenvalue of larger magnitude.  (c) A similar example for $\Lambda=-1.11$, $\Gamma=0.15$.}
    \label{fig:disorder}
\end{figure}

The inherent disorder in equilibrium monolayers is illustrated in Figure~\ref{fig:disorder}.  The variance of $P^{\text{eff}}_\alpha$ (about mean zero) within a monolayer of 800 cells is mapped at discrete locations across $(\Lambda,\Gamma)$-parameter space in Figure~\ref{fig:disorder}(a).  For each simulation, $\mathcal{L}$ was incrementally adjusted to enforce (\ref{eq:zeroload}).  The variability weakens near the region I/II$_a$ boundary and increases with $\Gamma$.  Two individual realisations  (Figure~\ref{fig:disorder}b,c) reveal mesoscopic patterns that emerge across the monolayer: shading identifies cells with positive or negative $P^{\text{eff}}_\alpha$ and line segments characterise the orientation of cell shape and stress.  The example closer to $\Lambda=0$ (Figure~\ref{fig:disorder}b) reveals slender patterns that are correlated over many cell lengths.  Cells that are larger (smaller) than their equilibrium area, with $P^{\text{eff}}>0$ ($<0$), tend to align with their principal axis of shape (and stress) parallel (perpendicular) to the line of cells, in structures that are reminiscent of force chains in jammed systems \citep{Majmudar2005}.  In particular, chains of darker cells are elongated parallel to the chain and exert a net tensile force along each chain, whereas lighter cells are compressed along their chain axis and exert a net compressive force along each chain.  {\color{black}Further visualisation of these structures is provided in Appendix~\ref{sec:appd} (Figure~\ref{fig:force_chains}a).}  In contrast, nearer the Region I boundary (Figure~\ref{fig:disorder}c), the correlation length of patterns increases and there appears to be less alignment of neighbouring cells.

\begin{figure}
	\centering
	\includegraphics[width=0.9\textwidth]{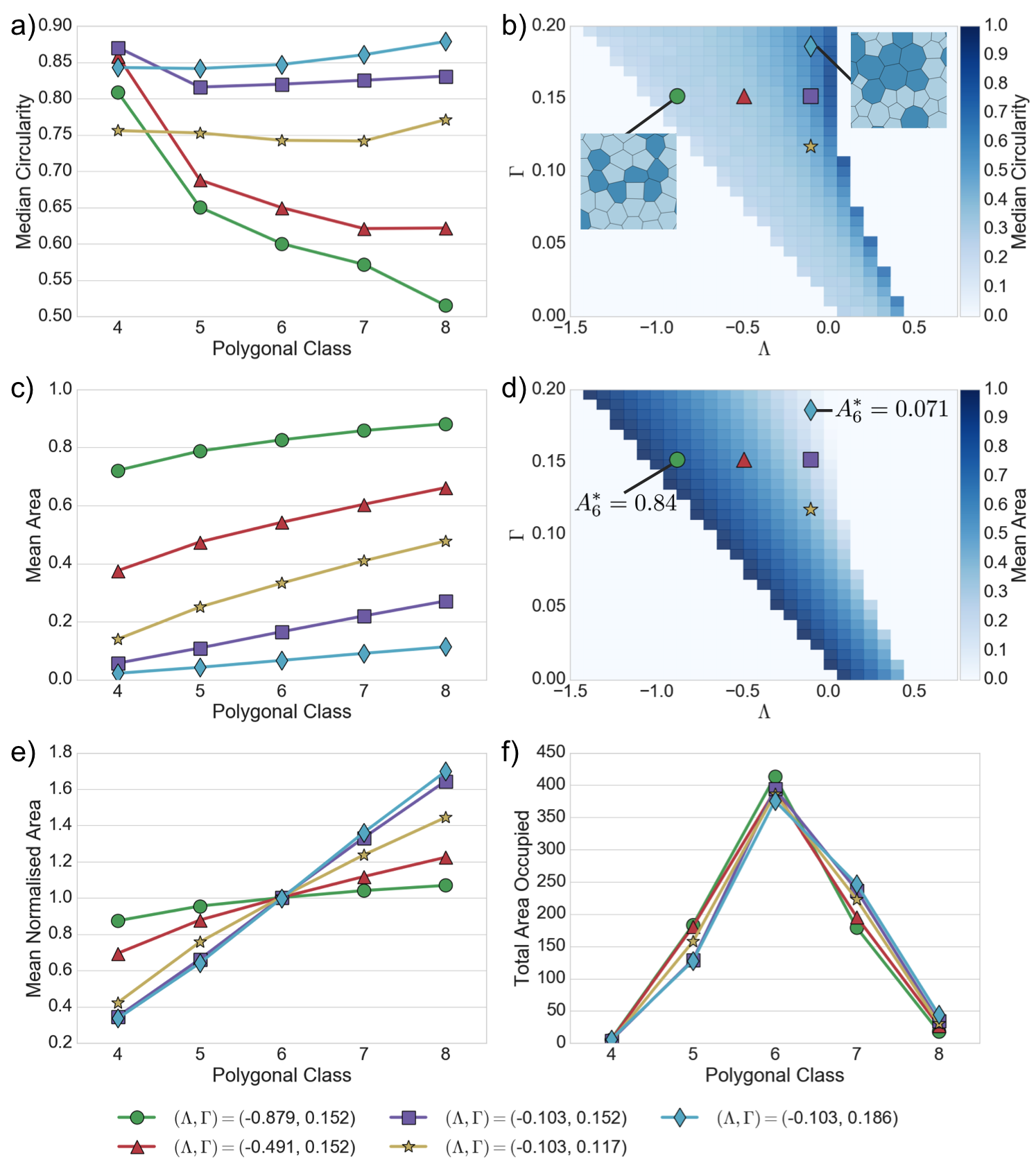}
	\caption{Dependence of cell geometry on model parameters, using \textcolor{black}{5 unique} simulations with 800 cells \textcolor{black}{(4000 cells total)} in a periodic box under zero net external pressure.  (a) Mean circularity of cells per polygonal class, at parameter values indicated by corresponding symbols in $(\Lambda,\Gamma)$-parameter space in (b,c).  (b) The heat map shows mean circularity of all cells in a simulation, using the same realisations used in Figure \ref{fig:disorder}.  Insets show two example configurations.  (c) Mean cell area per polygonal classes, for the same set of parameters. (d) Heat map of mean area of all cells across $(\Lambda,\Gamma)$-parameter space.  (e) Mean cell area per polygonal class for given parameters, normalised by the mean area of hexagons. (f) Total area of all cells in each polygonal class, such that the sum of all points equals the area of the box.}
    \label{fig:mean_area_circularity_poly_class}
\end{figure}

Figure \ref{fig:mean_area_circularity_poly_class} illustrates the impact of varying parameters $(\Lambda,\Gamma)$ (with $P_{\text{ext}}=0$) on the shape and size of cells when partitioned into polygonal classes.   The mean circularity of cells increases with $\Lambda$ as one moves across region II$_a$ (Figure \ref{fig:mean_area_circularity_poly_class}a,b): near the region-I boundary, cells with more sides become highly distorted (see inset), whereas near the region IIa/III boundary (where $L_0\rightarrow 0$) cells become more uniformly round.  Increasing $\Gamma$ for fixed $\Lambda$ near this boundary increases the cortical tension and promotes rounding, while reducing the mean cell area (Figure \ref{fig:mean_area_circularity_poly_class}c,d).  Moving back across region II$_a$ towards the region-I boundary, $L_0$ increases, reducing cortical tension and allowing cells to enlarge.  In comparison to the size of hexagons, the area distribution across polygonal classes (Figure~\ref{fig:mean_area_circularity_poly_class}e) is much more uniform near the region I/II$_a$ border than near the II$_a$/III border.  The non-linearity in $P^{\text{eff}}_N$ implies that changes in parameters influence circularity and areas among different polygonal classes non-uniformly.  In contrast, the total area occupied by different polygonal classes shows surprisingly little parameter variation (Figure~\ref{fig:mean_area_circularity_poly_class}f).

\begin{figure}
	\centering
	\includegraphics[width=.95\textwidth]{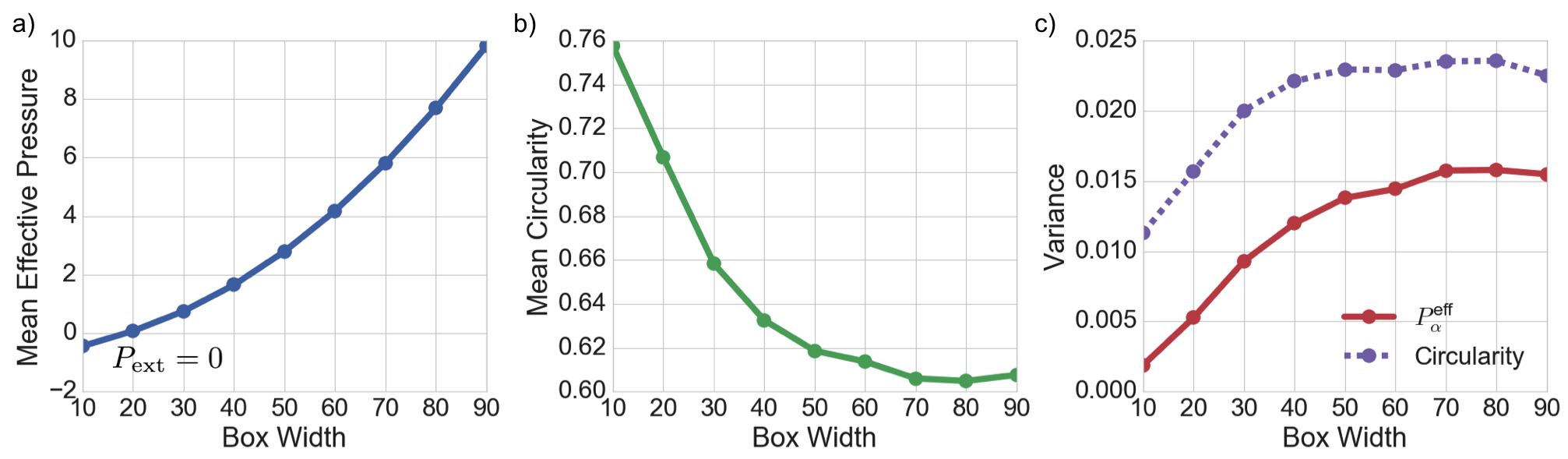}
		\caption{Visualising the effect of peripheral stress on network packing geometry.  800 cells were simulated in boxes of width $\mathcal{L}=10, 20, \dots, 90$ leading to $P^{\text{eff}}$ distributions with means shown in (a).  {\color{black}$P_{\text{ext}}=0$ for a box width of 20.} The corresponding means of the distributions of circularities are shown in (b). The variance of the distributions at different box widths are given in (c), for $P_\alpha^{\text{eff}}$ (solid) and circularity (dashed).  Model parameters used were $(\Lambda,\Gamma)=(-0.1,0.1)$ for which $A_6^*=0.446$.  Larger box sizes have lower cell density, higher mean $P^{\text{eff}}$, lower mean circularity and greater variability.}
    \label{fig:density_plots}
\end{figure}

In addition to the model parameters $(\Lambda, \Gamma)$, the density of cells (controlled by $P_{\text{ext}}$ in (\ref{eq:weight})) also induces changes in the equilibrium cell packing configurations.  As Figure~\ref{fig:density_plots} illustrates, monolayers under uniform net compression (for which $P_{\text{ext}} < 0$ on average) will tend to produce more round cells, closer to perfect polygons.   In contrast, monolayers under uniform net tension (for which $P_{\text{ext}} > 0$ on average) exhibit more disordered arrays, with cells tending to be more elongated.  In parameter fitting below, we initially impose the constraint $P_{\text{ext}}=0$.  

\subsection{4.1 Parameter fitting}

Of the features described in Figure~\ref{fig:mean_area_circularity_poly_class}, the total area per polygonal class (panel f) is a poor candidate for parameter identification, while the mean area (panel c) requires a dimensional measure of area and the mean normalised area (panel d) shows limited variation.  In contrast, the mean circularity (panel a) shows strong parameter variation without the additional requirement of a lengthscale measurement. {\color{black} However, searching across parameter space we found it difficult to capture simultaneously both the distribution of mean area and the distribution of mean cell circularity.   
Given the key contribution of cell area to the stress tensor, we therefore chose to use cell area (following \cite{Farhadifar:2007})} to parameterise the model to the \textit{Xenopus laevis} animal cap explants introduced in Section~\ref{sec:experiments}; {\color{black}we return to circularity below}.  

\begin{figure}
	\centering
	\includegraphics[width=0.9\textwidth]{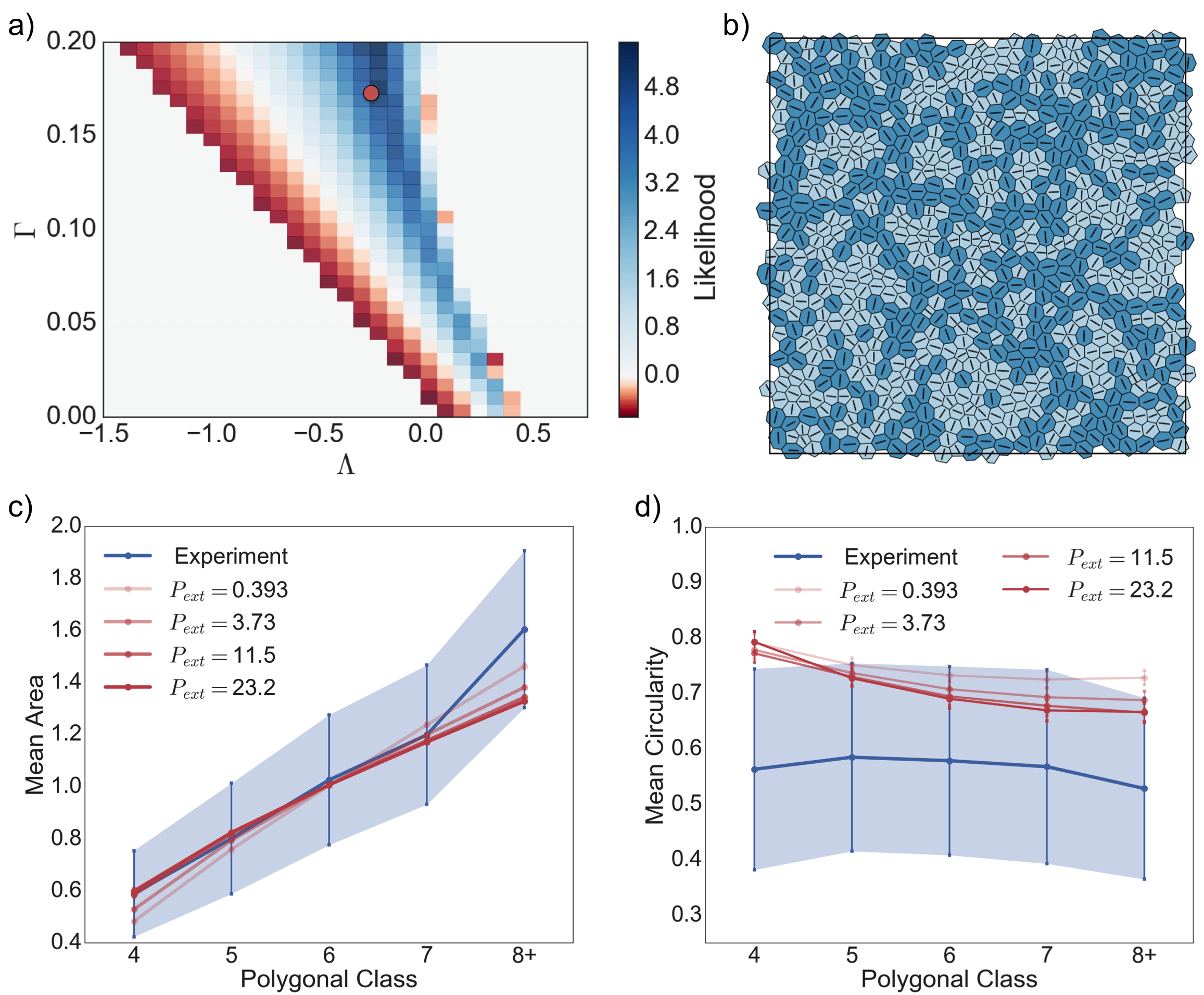}
	\caption{Results of parameter fitting. (a) Heat map showing value of the likelihood function \eqref{eq:log-posterior} across a uniform grid in valid parameter space. The simulated monolayers used were the same as those in Figures \ref{fig:disorder} and \ref{fig:mean_area_circularity_poly_class}. For each monolayer, the mean 
    \textcolor{black}{areas} per polygonal class were calculated and used to evaluate \eqref{eq:log-posterior}. The likelihood was maximised at $(\Lambda, \Gamma)\approx (-0.26, 0.17)$, marked by the circular symbol; a corresponding monolayer is shown in (b), with cells having $P_\alpha^{\text{eff}}>0$ ($<0$) shaded dark (light). {\color{black} (c, d) Distributions of area and circularity across polygonal classes for simulations with $(\Lambda^\dag,\Gamma^\dag)=(-0.259,0.172)$ for increasing values of $P_{\mathrm{ext}}$.}}
    \label{fig:best_param_fits}
\end{figure}

Using simulations of monolayers under  $P_{\text{ext}}=0$, we generated datasets {\color{black}$\boldsymbol{A}^{\;\text{sim}}(\Lambda,\Gamma)$, the mean areas} of cells in each polygonal class, to compare with experimental data {\color{black}$\mathbf{A}^{\;\text{exp}} = \{ \bar{A}_{4}^{\;\text{exp}}, \bar{A}_{5}^{\;\text{exp}}, \bar{A}_{6}^{\;\text{exp}}, \bar{A}_{7}^{\;\text{exp}}, \bar{A}_{8+}^{\;\text{exp}} \} $}.  We asses the fit of {\color{black}$\boldsymbol{A}^{\;\text{sim}}$} relative to {\color{black}$\boldsymbol{A}^{\;\text{exp}}$} using the following log-likelihood
{\color{black}
\begin{equation}
	\label{eq:log-posterior}
	\ln \left( \mathcal{P}(\Lambda,\Gamma \; | \; \mathbf{A}^{\;\text{exp}}, \mathbf{A}^{\;\text{sim}}) \right) \propto - \ln\left( \sum\limits_{i=4}^{8} |\bar{A}_{i}^{\;\text{exp}} - \bar{A}_{i}^{\;\text{sim}}(\boldsymbol{\theta})|^{2} \right).
\end{equation}
}
Evaluating (\ref{eq:log-posterior}) across a grid of parameter samples in region II (Figure~\ref{fig:best_param_fits}a), the posterior was maximised with {\color{black}$(\Lambda, \Gamma)\approx (-0.26, 0.17)$}, for which {\color{black}$L_0\approx 0.76$}.  While there are other credible parameter regions near the region III boundary, we can be confident that the monolayer in this experiment is far from the rigidity transition at region I, and reasonably certain that it falls outside region II$_b$ (where $L_0<0$).  The distribution of {\color{black}area} across polygonal classes is captured well by the model (Figure~\ref{fig:experiments}e).  For best-fit parameters, cells which are larger than average (shaded dark in Figure~\ref{fig:best_param_fits}b) tend to align in slender structures or, in some instances, to be isolated at the centre of a rosette of smaller (pale) cells.  

{\color{black} Despite matching area distributions well, the circularity distribution is  over-estimated across all polygonal classes (Figure~\ref{fig:experiments}f).  Figure~\ref{fig:density_plots}(b) suggests that the circularity can be reduced by putting the monolayer under net tension.  To investigate the possibility that the thin basal tissue layer of the animal cap (Figure~\ref{fig:experiments}a,b) might induce such a tension in the apical epithelium, we ran additional simulations for which $P_{\mathrm{ext}}>0$ (see (\ref{eq:weight})), maintaining fixed values of $\Lambda^\dag$ and $\Gamma^\dag$ (see (\ref{eq:trans})) in order to remain in an equivalent region of parameter space.  A demonstration of the changes in cell area and circularity across polygonal classes as $P_{\text{ext}}$ for $(\Lambda^{\dag},\Gamma^{\dag})=(-0.259,0.172)$ is given in Figure~\ref{fig:best_param_fits}(c,d).  While the area distribution maintains close agreement with experiment as $P_{\mathrm{ext}}$ increases, the circularity moves towards the experimental range but does not fall comfortably within it, even for very large $P_{\mathrm{ext}}$.  We conclude that additional refinements to the model (such as higher order nonlinearities in the energy $\tilde{U}_\alpha$, see (\ref{eq:single_cell_energy})) may be necessary to ensure quantitative agreement of both area and circularity distributions.  }

\section{Discussion}

We have investigated a popular vertex-based model of planar epithelia, addressing features associated with cell packing rather than division or motility.  We focused on a simple version of the model, neglecting refinements such as representations of internal viscous forces \citep{okuda2015}, non-planarity {\color{black}\citep{bielmeier2016, hannezo2014, Murisic:2015}}, descriptions of curved cell edges {\color{black}\citep{brodland2014, ishimoto2014}}, internal anisotropy, multiple cell types and so on. We first derived an expression (\ref{eq:full_stress_inc_dissipative}) for the stress $\boldsymbol{\sigma}_\alpha$ of an individual cell, expressed in terms of its shape.  The isotropic component of stress reveals the cell's effective pressure $P^{\text{eff}}_\alpha$ \eqref{eq:peff}, which is set by a balance between the internal pressure associated with bulk (cytoplasmic) forces that regulate cell area and cortical tension that regulates the cell perimeter.  With the area below and the perimeter above their respective targets ($A_\alpha<1$ and $L_\alpha>L_0$), the bulk forces push outward against the stretched perimeter, giving the cell some rigidity.   The traceless tensor $\mathsf{J}_\alpha$ in (\ref{eq:full_stress_inc_dissipative}) characterises asymmetries in the cell shape that might arise from an imposed shear stress or, in the absence of an external load, internal asymmetries associated with intrinsic disorder.  A simple representation of viscous forces associated with drag from the underlying substrate leads to a further contribution to the stress associated with dynamic shape changes.  Crucially, the principal axes of the shape tensor $\mathsf{S}_\alpha$ (defined in terms of the vertex locations) align {\color{black}exactly} with the principal axes of the cell stress, as illustrated in Figure~\ref{fig:stress_shape_align}.  This result may have implications in cell division, where it is postulated that there may be shape- and stress-sensing mechanisms guiding the positioning of the mitotic spindle \citep{Minc:2011, thery:2006}.  If the vertex-based model is accepted as a leading-order description of cell mechanics, it follows that it will not be possible to separate these mechanisms by looking solely at cell geometry, since the orientation of any inferred stress will necessarily align with the cell shape.  Instead, the system must be perturbed, either mechanically or chemically (using biological knockdowns, for example), such that the mechanisms can be disrupted and separated.  {\color{black}In this context, it is worth highlighting the distinction between the orientation of external stress that may be imposed on a monolayer, and the heterogeneous stress field at the individual cell level (\hbox{e.g.} Figure~\ref{fig:stress_shape_align}).  Observations show cell division in a stretched monolayer to be aligned with cell shape rather than the external stress orientation \citep{Wyatt2015}; the present model suggests that the cell-scale stress would be aligned with cell shape, even if the average stress at monolayer level has a different orientation.}

{\color{black} The distinction between individual cell stress $\boldsymbol{\sigma}_\alpha$ and tissue-level stress $\boldsymbol{\sigma}^R$ is evident in the} expression (\ref{eq:tissuestr}) for the stress over a patch of cells, derived as an area-weighted average of the individual cell stresses.  For a monolayer under an isotropic external load of magnitude $P_{\text{ext}}$, we derived a constraint (\ref{eq:weight}) on the area-weighted $P^{\text{eff}}_\alpha$; furthermore, the averaged deviatoric stress must vanish in this case.  When simulating a monolayer that is not subject to lateral forcing, the constraint of zero mean effective pressure (\ref{eq:zeroload}) is important in determining the appropriate cell density within the simulation domain.   One can then examine the properties of the monolayer when this configuration is perturbed by small compressive or shear deformations.  We derived an exact expression (\ref{eq:bulk}) for the monolayer's bulk elastic modulus (generalising results obtained previously for hexagonal cell arrays) and recovered directly an expression (\ref{eq:shear_hex}) for the shear modulus in the hexagonal packing limit.  The mechanical properties of the tissue can therefore be tuned by varying the relative strengths of the bulk and cortical forces.  As shown previously \citep{bi2015}, a phase transition arises when $L_0\approx 3.81$, which bounds a region of parameter space in which the monolayer loses resistance to shear deformations.   Fitting our model to data from embryonic \textit{Xenopus laevis} tissue, by maximising a likelihood function derived from the mean {\color{black}area} per polygonal class, suggests {\color{black}$L_0\approx 0.76$} in the embryonic tissue, substantially distant from the rigidity transition.  {\color{black} The model fit is imperfect however, as we were not able to capture circularity distributions even when varying the peripheral load on the monolayer (Figures~\ref{fig:experiments}f, \ref{fig:best_param_fits}). This suggests further constitutive refinements of the model are needed, such as including higher-order nonlinearities in \eqref{eq:single_cell_energy}.}  We also examined how cell shape (and of course size) can be influenced by an external load $P_{\text{ext}}$, with cells becoming rounder when tightly packed (Figure~\ref{fig:density_plots}).  The bulk isotropic stress (or equivalently the mean cell density) is likely to be a significant parameter when simulating confined tissues, and is an example of a mechanical signal that can be communicated over long distances.  {\color{black}Future studies should address anisotropic external loading, which has the capacity to promote more ordered cell packing \citep{sugimura2013}.}

The present descriptions of the stress tensor are appropriate for small-amplitude deformations close to equilibria, and in future should be extended to account for irreversible cell rearrangements (such as T1/T2 transitions) that endow the material with an elastic-viscoplastic character, as well as accounting for cell division.  Kinematic and geometric quantities (such as the texture tensor) characterising large deformations of cellular materials have been developed that are based on connections between centres of adjacent cells {\color{black}\citep{blanchard2009, blanchard2017, etournay2015, graner2008, guirao2015, tlili2015}}, the dual network to that considered here.  While it is straightforward to repartition the stress (\ref{eq:tissuestr}) over the network of triangles connecting cell centres, it is less clear how to relate it to strain measures defined with respect to cell centres rather than cell vertices, {\color{black}without for example assuming that vertices are barycentric with respect to cell centres \citep{barton2016}}.  In particular, the relationship between the tissue-level stress postulated by \cite{etournay2015} to that emerging from the vertex-based model remains to be established.

While the monolayer can be stress-free at the bulk scale, individual cells can have non-zero $P^{\text{eff}}_\alpha$: those for which $P^{\text{eff}}_\alpha>0$ ($<0$) are larger (smaller) than the equilibrium area at which bulk and cortical forces balance.   Each simulation of a spatially disordered monolayer describes an equilibrium configuration of this very high-dimensional dynamical system, subject to the constraint that all edge lengths exceed a defined threshold (smaller edges being removed by T1/T2 transitions).  We have characterised some features of the variability of these states, both in terms of the variance in $P^{\text{eff}}_\alpha$ over the cell population and the spatial pattern of compressed and dilated cells.   While soft monolayers near the region I/II boundary show very long-range patterning (Figure~\ref{fig:disorder}c), stiffer monolayers nearer the II/III boundary {\color{black}appear to} exhibit chains of force (and cell shape, Figures~\ref{fig:disorder}b, {\color{black}\ref{fig:force_chains}a}), where lines of tension and compression are transmitted along entangled strings.  Evidence of force chains has recently been provided in the \emph{Drosophila melanogaster} embryo \citep{JasonGao2016} and the patterns suggested by our model (Figure~\ref{fig:best_param_fits}) motivate ongoing investigations in the \emph{Xenopus} system.  Robust evidence of force-shape chains in real epithelia would raise interesting questions about the role of mechanical feedback on patterning of cell division.

\section*{Acknowledgements} 

ANB was supported by a BBSRC studentship.  OEJ acknowledges EPSRC grant EP/K037145/1. SW and GG are supported by a Wellcome Trust/Royal Society Sir Henry Dale Fellowship to SW [098390/Z/12/Z].

 \bibliographystyle{plainnat}
\bibliography{bib}

\begin{appendix}

\section{Geometric identities}

The contribution to the stress due to cell pressure first involves
\begin{equation}
    \sum_{i=0}^{Z_{\alpha}-1} \mathbf{R}_{\alpha}^i\otimes \mathbf{p}_{\alpha}^i =  \tfrac{1}{2} \sum_{i=0}^{Z_{\alpha}-1} \mathbf{R}_{\alpha}^i\otimes \left[ \left( \mathbf{R}_{\alpha}^{i+1} - \mathbf{R}_{\alpha}^{i-1} \right) \times \mathbf{\hat{z}} \right].
\end{equation}
    Taking components, 
\begin{equation}
    \begin{split}
    \left\{ \sum_{i=0}^{Z_{\alpha}-1} \mathbf{R}_{\alpha}^i\otimes \mathbf{p}_{\alpha}^i \right\}_{pq} 
    &=\tfrac{1 }{2}\sum_{i=0}^{Z_{\alpha}-1}R_{\alpha,p}^i \left[\delta_{q1}     (R_{\alpha,2}^{i+1}-R_{\alpha,2}^{i-1})-\delta_{q2} (R_{\alpha,1}^{i+1}-R_{\alpha,1}^{i-1})\right] = A_{\alpha} \delta_{pq} 
    \end{split}
\end{equation}
giving (\ref{eq:geomid}a).
Referring now to the contractility term, we find
\begin{equation}
    \begin{split}
    \sum_{i=0}^{Z_{\alpha}-1} \mathbf{R}_{\alpha}^i\otimes \mathbf{q}_{\alpha}^i &= \sum_{i=0}^{Z_{\alpha}-1} \mathbf{R}_{\alpha}^i\otimes (\hat{\mathbf{t}}_{\alpha}^i-\hat{\mathbf{t}}_{\alpha}^{i-1}) 
     = \sum_{i=0}^{Z_{\alpha}-1} \mathbf{R}_{\alpha}^i\otimes \left[ \frac{\mathbf{t}_{\alpha}^i}{l_\alpha^i}
 - \frac{\mathbf{t}_{\alpha}^{i-1}}{l_\alpha^{i-1}} \right].
    \end{split}
\end{equation}
recalling $l_{\alpha}^{i} = (\mathbf{t}_{\alpha}^i \cdot \mathbf{t}_{\alpha}^i)^{\frac{1}{2}}$.  Thus
\begin{equation}
    \label{eq:J}
    \begin{split}
    \sum_{i=0}^{Z_{\alpha}-1} \mathbf{R}_{\alpha}^i\otimes \mathbf{q}_{\alpha}^i &= \sum_{i=0}^{Z_{\alpha}-1} \mathbf{R}_{\alpha}^i\otimes \left[ (l_{\alpha}^{i})^{-1} \mathbf{R}_{\alpha}^{i+1} + (l_{\alpha}^{i-1})^{-1}\mathbf{R}_{\alpha}^{i-1} - ((l_{\alpha}^{i})^{-1} + (l_{\alpha}^{i-1})^{-1}) \mathbf{R}_{\alpha}^{i} \right] \\
    &=  \sum_{i=0}^{Z_{\alpha}-1} \mathbf{R}_{\alpha}^{i} \otimes ((l_{\alpha}^{i})^{-1}\mathbf{R}_{\alpha}^{i+1} + (l_{\alpha}^{i-1})^{-1}\mathbf{R}_{\alpha}^{i-1} ) -  ((l_{\alpha}^{i})^{-1} + (l_{\alpha}^{i-1})^{-1})\mathbf{R}_{\alpha}^{i} \otimes \mathbf{R}_{\alpha}^{i} \\
    & = \sum_{i=0}^{Z_{\alpha}-1} (l_{\alpha}^{i})^{-1} \left[ \mathbf{R}_{\alpha}^{i} \otimes (\mathbf{R}_{\alpha}^{i+1} - \mathbf{R}_{\alpha}^{i}) + \mathbf{R}_{\alpha}^{i+1} \otimes (\mathbf{R}_{\alpha}^{i} - \mathbf{R}_{\alpha}^{i+1}) \right]  \\
    & =  \sum_{i=0}^{Z_{\alpha}-1} (l_{\alpha}^{i})^{-1} \left[ \mathbf{R}_{\alpha}^{i} \otimes \mathbf{t}_{\alpha}^{i} - \mathbf{R}_{\alpha}^{i+1} \otimes \mathbf{t}_{\alpha}^{i} \right]  \\
	&= \sum_{i=0}^{Z_{\alpha}-1} (l_{\alpha}^{i})^{-1} \left[( \mathbf{R}_{\alpha}^{i} - \mathbf{R}_{\alpha}^{i+1} ) \otimes \mathbf{t}_{\alpha}^{i} \right] = - \sum_{i=0}^{Z_{\alpha}-1} \hat{\mathbf{t}}_{\alpha}^{i} \otimes \mathbf{t}_{\alpha}^{i}
    \end{split}
\end{equation}
giving (\ref{eq:geomid}b).  This is symmetric because $\hat{\mathbf{t}}_{\alpha}^{i} \otimes \mathbf{t}_{\alpha}^{i}={\mathbf{t}}_{\alpha}^{i} \otimes \hat{\mathbf{t}}_{\alpha}^{i}$. 

\section{Proof that $\mathsf{S}_\alpha$, {\color{black}$\dot{\mathsf{S}}_\alpha$} and $\mathsf{J}_\alpha$ align} 
\label{sec:s_j_align}

To establish that $\mathsf{S}_\alpha\mathsf{J}_\alpha =  \mathsf{J}_\alpha \mathsf{S}_\alpha$ for cell $\alpha$, we can ignore the pre-factors in the tensors and need only show
\begin{equation}
    \label{eq:sj_js}
	\left( \sum_{i=0}^{Z_{\alpha}-1} \mathbf{R}_{\alpha,p}^{i}\mathbf{R}_{\alpha,q}^{i} \right)\left( \sum_{j=0}^{Z_{\alpha}-1} \hat{\mathbf{t}}_{\alpha,q}^{j}\mathbf{t}_{\alpha,r}^{j} \right) = \left( \sum_{i=0}^{Z_{\alpha}-1} \hat{\mathbf{t}}_{\alpha,p}^{i}\mathbf{t}_{\alpha,q}^{i} \right)\left( \sum_{j=0}^{Z_{\alpha}-1} \mathbf{R}_{\alpha,q}^{j}\mathbf{R}_{\alpha,r}^{j} \right)
\end{equation}
Let us henceforth assume that the sums over $i,j$ and $q$ are implicit. We also drop the $\alpha$ subscripts, under the assumption that all vectors are relative to the same cell centroid. Considering the left hand side (LHS) first:
\begin{equation}
\label{eq:lhs}
	\begin{split}
	\text{LHS} &= R_{p}^{i}R_{q}^{i}\left( R_{q}^{j+1} - R_{q}^{j} \right)\left( R_{r}^{j+1} - R_{r}^{j} \right) / l^{j} \\
    &= R_{p}^{i}R_{q}^{i}\left( R_{q}^{j+1}R_{r}^{j+1} + R_{q}^{j}R_{r}^{j} \right) / l^{j} - R_{p}^{i}R_{q}^{i}\left( R_{q}^{j+1}R_{r}^{j} + R_{q}^{j}R_{r}^{j+1} \right) / l^{j} \\
	&= M_{pq}^{i}M_{qr}^{j} /(2 l^{j}) - M_{pq}^{i}N_{qr}^{j} /(2 l^{j}) \equiv \tfrac{1}{2}( \barroman{I}_{pr} - \barroman{II}_{pr})
	\end{split}
\end{equation}
where $M_{pq}^{i} = 2 R^{i}_{p}R^{i}_{q} = R_{p}^{i}R_{q}^{i} + R_{p}^{i+1}R_{q}^{i+1}$ and $N_{qr}^{j} = R_{q}^{j+1}R_{r}^{j} + R_{q}^{j}R_{r}^{j+1}$ are symmetric ($M_{pq}^{i} = M_{qp}^{i}$ and $N_{qr}^{j} = N_{rq}^{j}$) and
\begin{equation}
	\begin{split}
	\barroman{I}_{pr} &\equiv M_{pq}^{i}M_{qr}^{j} / l^{j} = M_{qp}^{i}M_{rq}^{j} / l^{j} = M_{qp}^{j}M_{rq}^{i} / l^{i} = (M_{rq}^{i} / l^{i}) M_{qp}^{j} \\
	&= (M_{pq}^{i} / l^{i}) M_{qr}^{j}
	\end{split}
\end{equation}
where we have exchanged indices in the first line and made use of the symmetry of the product in the second. By similar steps we find
\begin{equation}
	\begin{split}
	\barroman{II}_{pr} &\equiv M_{pq}^{i}N_{qr}^{j} / l^{j} = M_{qp}^{i}N_{rq}^{j} / l^{j} = M_{qp}^{j}N_{rq}^{i} / l^{i} = (N_{rq}^{i} / l^{i}) M_{qp}^{j} \\
	&= (N_{pq}^{i} / l^{i}) M_{qr}^{j}.
	\end{split}
\end{equation}
However, noting the definitions above, we see that the right hand side (RHS) of \eqref{eq:sj_js} may be written as
\begin{equation}
	\begin{split}
	\text{RHS} &= (M_{pq}^{i} / 2l^{i}) M_{qr}^{j} - (N_{pq}^{i} / 2l^{i}) M_{qr}^{j} =\tfrac{1}{2}( \barroman{I}_{pr} - \barroman{II}_{pr} )
	\end{split}
\end{equation}
matching (\ref{eq:lhs}).  Therefore the tensors commute and we have alignment of the principal axes of stress and shape, when the system is in equilibrium. 

{\color{black} Let us now establish $\dot{\mathsf{S}}_\alpha\mathsf{S}_\alpha =  \mathsf{S}_\alpha \dot{\mathsf{S}}_\alpha$. Ignoring pre-factors again, we have
\begin{equation}
    \begin{split}
    \dot{\mathsf{S}}_\alpha\mathsf{S}_\alpha &= \frac{1}{2}( \dot{R}_{p}^{i}R_{q}^{i} + R_{p}^{i}\dot{R}_{q}^{i} )R_{q}^{j}R_{r}^{j}  = \dot{R}_{p}^{i}R_{q}^{i}R_{q}^{j}R_{r}^{j},
    \end{split}
\end{equation}
which is symmetric. Given that $\dot{\mathsf{S}}_\alpha$, $\mathsf{S}_\alpha$ are both symmetric, and their product is symmetric, we have a necessary and sufficient condition that they commute.  We therefore also have alignment of the principal axes of stress and shape when the system is out of equilibrium. }

\section{Shear modulus of a perfectly hexagonal cell}
\label{sec:shear}

For a 2D linearly elastic isotropic material with constitutive relation $\boldsymbol{\sigma}=K\mathsf{I}\mathrm{Tr}(\boldsymbol{\varepsilon})+2G(\boldsymbol{\varepsilon}-\tfrac{1}{2}\mathsf{I}\mathrm{Tr}(\boldsymbol{\varepsilon}))$, where $K$ is bulk modulus, $G$ shear modulus and $\boldsymbol{\varepsilon}$ linear strain, a small shear deformation $(x_1,x_2)=(X_1,X_2)+\gamma (X_2,0)$ (defined with respect to Cartesian axes mapping $\mathbf{X}$ to $\mathbf{x}$), with $\boldsymbol{\varepsilon}=\tfrac{1}{2}\gamma(\hat{\mathbf{x}}_1\otimes \hat{\mathbf{x}}_2+\hat{\mathbf{x}}_2\otimes \hat{\mathbf{x}}_1)$, generates a shear stress $\sigma_{12}=\gamma G$.  We expect a cell array formed from perfect hexagons to be characterised by effective isotropic material parameters $K$ and $G$.  We discard the subscript $\alpha$ and let a representative cell have vertices $\mathbf{R}^i=(L/6)\mathbf{r}^i$ where $\mathbf{r}^i=(c_i,s_i)$, $c_i\equiv \cos(\pi i/3)$, $s_i=\sin(\pi i/3)$ and $A = (L / \mu_{6})^{2}$.  We can then identify $G$ by perturbing the equilibrium stress $\boldsymbol{\sigma}(\mathbf{R}^i)=-P^{\mathrm{eff}}\mathsf{I}+T\mathsf{J}$ under the given shear deformation and Taylor expanding $\boldsymbol{\sigma}(\mathbf{R}^i+\gamma(R_2^i,0))$ about the equilibrium state for which $P^{\mathrm{eff}}=0$ and $\mathsf{J}=\mathsf{0}$.  Thus we must evaluate
\begin{equation} 
G=-\sum_{i=0}^{5} R_2^i \frac{\partial \sigma_{12}}{\partial R_1^i} =  \sum_{i=0}^{5} R_2^i \frac{\partial }{\partial R_1^i} \left( \frac{\Gamma (L-L_0)}{ A} \sum_{k=0}^{5} \frac{{t}^k_1 {t}^k_2}{l^k} \right)
\end{equation}
in the symmetric configuration. The initial minus sign arises because $\boldsymbol{\sigma}$ models the restoring cell forces, whereas the shear modulus is calculated using the force required to deform the object. The sum over $i$ arises from the chain rule. The sum over $k$ vanishes in the equilibrium configuration so we need consider only its derivatives, for which the only nonzero contributions are when $k=i-1$ and $k=i$. Performing the differentiation, we have
\begin{align}
    G &= \frac{\Gamma (L-L_0)}{ A} \sum_{i=0}^{5} R_2^i \sum_{k=0}^{5} \frac{1}{(l^{k})^{2}} \left( l^{k} \frac{\partial (t^{k}_{1}t^{k}_{2})}{\partial R^{i}_{1}} - \frac{\partial l^{k}}{\partial R^{i}_{1}} (t^{k}_{1}t^{k}_{2}) \right),
\end{align}
where 
\begin{equation}
    \sum_{k=0}^{5} \frac{1}{(l^{k})^{2}} \left( l^{k} \frac{\partial (t^{k}_{1}t^{k}_{2})}{\partial R^{i}_{1}} \right) = ( 2R^i_2 - R^{i+1}_2 - R^{i-1}_2)/l^i
    =  2s_{i} - s_{i+1} - s_{i-1}
\end{equation}
and
\begin{align}
	\sum_{k=0}^{5} \frac{1}{(l^{k})^{2}} \left( \frac{\partial l^{k}}{\partial R^{i}_{1}} (t^{i}_{1}t^{i}_{2}) \right) 
	& = \left((R^{i}_1 - R^{i-1}_1)\hat{t}^{i-1}_{1}\hat{t}^{i-1}_{2} - (R^{i+1}_1 - R^{i}_1)\hat{t}^{i}_{1}\hat{t}^{i}_{2} \right)/l^i \\
	&= (c_{i} - c_{i-1})\hat{t}^{i-1}_{1}\hat{t}^{i-1}_{2} - (c_{i+1} - c_{i})\hat{t}^{i}_{1}\hat{t}^{i}_{2}
\end{align}
for ${\mathbf{t}}^{k}=l^{k}\hat{\mathbf{t}}^{k}$. Finally, evaluating the sum over $i$, we recover (\ref{eq:shear_hex}).

\begin{figure}
	\centering
	\includegraphics[width=0.95\textwidth]{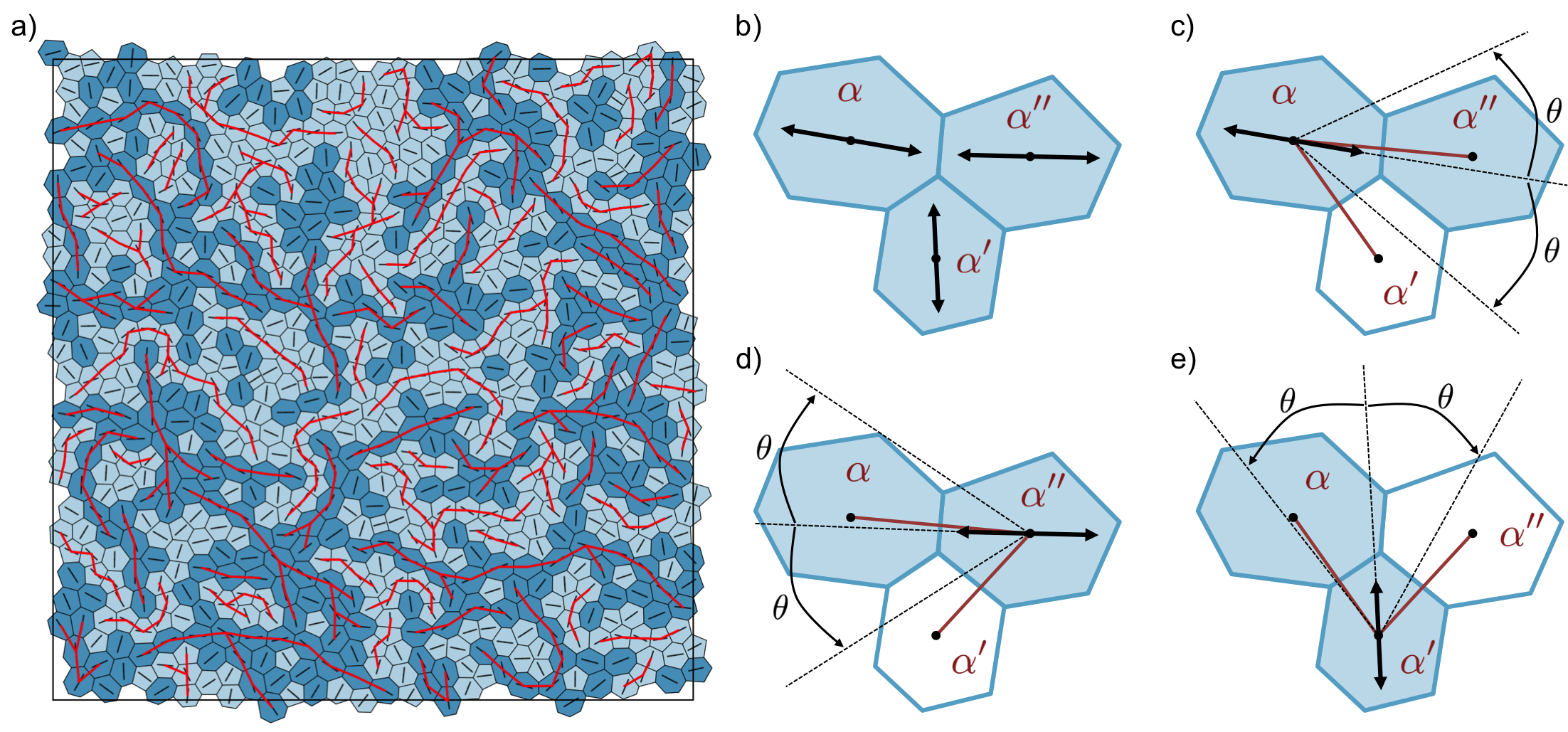} 
	\caption{{\color{black}(a)  An example of force chains in a monolayer, with 800 cells and $\Lambda=-0.1$, $\Gamma=0.1$, $P_{\mathrm{ext}=0}$.  Darker (lighter) shading denotes cells with $P^{\text{eff}}>0$ $(<0)$.  Short line segments indicate the principal axis of the stress tensor for each cell (see Figure~\ref{fig:ground_state}).  Long red lines identify chains satisfying (\ref{eq:chain_criteria}) with $\theta = \pi / 4$. (b-e) Illustration of criterion used to identify force chains.  Red lines represent vectors running between cell centroids.  Black double sided arrows indicate the principal axis of stress.  b) Cell $\alpha$ has been selected to start a chain, and cells $\alpha^{\prime}$ and $\alpha^{\prime\prime}$ are found to satisfy (\ref{eq:chain_criteria}c). (c-e)  Only $\alpha^{\prime\prime}$ is selected to join the chain as it satisfies both (\ref{eq:chain_criteria}a) (c) and (\ref{eq:chain_criteria}b) (d).  $\alpha^{\prime}$ is excluded because is fails (\ref{eq:chain_criteria}a) (c), despite satisfying (\ref{eq:chain_criteria}b) (e).}}
    \label{fig:force_chains}
\end{figure}

{\color{black}
\section{Visualising force chains}
\label{sec:appd}

 We identify force chains in the monolayers using a criterion adapted from \cite{Peters:2005}. In order for two cells, $\alpha$ and $\alpha^{\prime}$, to be in a force chain we require the following conditions to be satisfied:
    \begin{equation}
    \cos\theta < \frac{ \boldsymbol{\sigma}_{\alpha,1} \cdot (\mathbf{R}_{\alpha^{\prime}} - \mathbf{R}_{\alpha}) }{ \left|\left|\boldsymbol\sigma_{\alpha,1} \right|\right| \left |\left| \mathbf{R}_{\alpha^{\prime}} - \mathbf{R}_{\alpha}\right |\right | }, \quad
    \cos\theta < \frac{ \boldsymbol{\sigma}_{\alpha^{\prime},1} \cdot (\mathbf{R}_{\alpha} - \mathbf{R}_{\alpha^{\prime}}) }{ \left|\left|\boldsymbol\sigma_{\alpha^{\prime},1} \right|\right| \left |\left| \mathbf{R}_{\alpha} - \mathbf{R}_{\alpha^{\prime}}\right |\right | }, \quad
    0 < \sigma_{\alpha,1}  \sigma_{\alpha^{\prime},1}.
\label{eq:chain_criteria}
    \end{equation}
 Here $\sigma_{\alpha,1}$ ($\boldsymbol\sigma_{\alpha,1}$) is the principal eigenvalue (eigenvector) of the stress tensor of cell $\alpha$ and $\mathbf{R}_{\alpha^{\prime}} - \mathbf{R}_{\alpha}$ is the vector running from the centroid of cell $\alpha$ to the centroid of $\alpha^{\prime}$. Equation (\ref{eq:chain_criteria}a) ensures that cell $\alpha^{\prime}$ lies within $\theta$ radians of $\boldsymbol\sigma_{\alpha,1}$, while (\ref{eq:chain_criteria}b) equivalently ensures that cell $\alpha$ lies within $\theta$ radians of $\boldsymbol\sigma_{\alpha^{\prime},1}$.  (This reciprocal requirement  is demonstrated in Figure~\ref{fig:force_chains}(b-e); cell $\alpha$ lies within the criterion for cell $\alpha^{\prime}$, but $\alpha^{\prime}$ does not satisfy the criterion for $\alpha$, so the cells do not form a chain; however $\alpha$ and $\alpha''$ do form a chain.) Finally, (\ref{eq:chain_criteria}c) ensures that both cells are under compression or tension.

To construct the visualisation shown in Figure~\ref{fig:force_chains}(a), cells are randomly selected to start new chains, and this starting cell is then denoted a leader. Leaders are cells at the ends of chains, which have not had the above criterions checked with all of their neighbours. Once a new leader has been chosen to start a chain, the following procedure is executed:
\begin{enumerate}
    \item Select a leader from the current chain. This cell is no longer a leader.
    \item Identify all of this cell's neighbours which are not already part of a chain (including the current chain), if any. All neighbours that satisfy \eqref{eq:chain_criteria} are added to the chain and become new leaders.
    \item Repeat from step 1 until no leaders remain. 
\end{enumerate}
We chose to only include chains comprised of three or more cells. The fact that new leaders cannot neighbour current members of the chain ensures that we have no closed loops, although we do allow branching. However, it also means that the set of chains in a monolayer is not unique, but depends on which cells are chosen to start new chains. 
}

\end{appendix}

\end{document}